\documentclass[english, aps,prb,twocolumn,10pt, superscriptaddress,floatfix, citeautoscript]{revtex4-1}

\pdfoutput = 1
\usepackage[utf8]{inputenc}
\usepackage[T1]{fontenc}
\usepackage{units}
\usepackage{mathtools}
\usepackage{amsmath}
\usepackage{amssymb}
\usepackage{braket}
\usepackage{graphicx}
\usepackage{wasysym}
\usepackage{siunitx}
\usepackage{xcolor}
\usepackage[colorlinks, citecolor={blue!50!black}, urlcolor={blue!50!black}, linkcolor={red!50!black}]{hyperref}
\usepackage{tabularx}
\usepackage{multirow}
\usepackage{microtype}
\usepackage{babel}
\usepackage{layouts} 
\usepackage{bbm}

\DeclareMathOperator{\Tr}{Tr}
\DeclareMathOperator{\diag}{diag}

\begin{document}

\newcommand{\phosphor}{$^{31} \mathrm{P}$ }
\newcommand{\As}{$^{75} \mathrm{As} $ }
\newcommand{\Sblight}{$^{121} \mathrm{Sb}$ }
\newcommand{\Sb}{$^{123} \mathrm{Sb}$ }
\newcommand{\Bi}{$^{209} \mathrm{Bi}$ }
\newcommand{\Sipure}{$^{28} \mathrm{Si}$ }
\newcommand{\Sispin}{$^{29} \mathrm{Si}$ }

\title{Exploring quantum chaos with a single nuclear spin}

\author{Vincent Mourik}
\thanks{These authors contributed equally.}
\affiliation{Centre for Quantum Computation and Communication Technologies, School of
Electrical Engineering and Telecommunications, UNSW Sydney, Sydney, New
South Wales 2052, Australia}

\author{Serwan Asaad}
\thanks{These authors contributed equally.}
\affiliation{Centre for Quantum Computation and Communication Technologies, School of
Electrical Engineering and Telecommunications, UNSW Sydney, Sydney, New
South Wales 2052, Australia}

\author{Hannes Firgau}
\affiliation{Centre for Quantum Computation and Communication Technologies, School of
Electrical Engineering and Telecommunications, UNSW Sydney, Sydney, New
South Wales 2052, Australia}

\author{Jarryd J. Pla}
\affiliation{School of
Electrical Engineering and Telecommunications, UNSW Sydney, Sydney, New
South Wales 2052, Australia}

\author{Catherine Holmes}
\affiliation{School of Mathematics and Physics, The University of Queensland,
St Lucia, Brisbane 4072, Australia}

\author{Gerard J. Milburn}
\affiliation{ARC Centre of Excellence for Engineered Quantum Systems, School of Mathematics and Physics, The University of Queensland,
St Lucia, Brisbane 4072, Australia}

\author{Jeffrey C. McCallum}
\affiliation{Centre for Quantum Computation and Communication Technology, School of
Physics, University of Melbourne, Melbourne VIC 3010, Australia}

\author{Andrea Morello}
\affiliation{Centre for Quantum Computation and Communication Technologies, School of
Electrical Engineering and Telecommunications, UNSW Sydney, Sydney, New
South Wales 2052, Australia}

\date{\today}

\begin{abstract}
Most classical dynamical systems are chaotic. 
The trajectories of two identical systems prepared in infinitesimally different initial conditions diverge exponentially with time. 
Quantum systems, instead, exhibit quasi-periodicity due to their discrete spectrum. 
Nonetheless, the dynamics of quantum systems whose classical counterparts are chaotic are expected to show some features that resemble chaotic motion. 
Among the many controversial aspects of the quantum-classical boundary, the emergence of chaos remains among the least experimentally verified. 
Time-resolved observations of quantum chaotic dynamics are particularly rare, and as yet unachieved in a single particle, where the subtle interplay between chaos and quantum measurement could be explored at its deepest levels. 
We present here a realistic proposal to construct a chaotic driven top from the nuclear spin of a single donor atom in silicon, in the presence of a nuclear quadrupole interaction. 
This system is exquisitely measurable and controllable, and possesses extremely long intrinsic quantum coherence times, allowing for the observation of subtle dynamical behavior over extended periods. 
We show that signatures of chaos are expected to arise for experimentally realizable parameters of the system, allowing the study of the relation between quantum decoherence and classical chaos, and the observation of dynamical tunneling.
\end{abstract}

\maketitle

\section{Introduction} \label{sec: intro}

\subsection{Quantum chaos} \label{subsec: intro_background}

The correspondence principle, as formulated by the Copenhagen school of quantum mechanics, states that the dynamics of quantum systems should converge towards classical dynamics, in the limit where the system becomes large.
Appealing (and, for simple cases, often correct) as it may sound, this point of view is afflicted by a plethora of complications and controversies around the precise nature of the quantum-classical transition \cite{leggett2002}, such as decoherence \cite{zurek2003} and  the quantum measurement problem \cite{schlosshauer2005}. 
Another key aspect of the quantum-classical transition concerns reconciling the chaotic dynamics of certain classical systems with the unitary evolution of their quantum-mechanical counterparts. 

Classical chaos is ubiquitous and well understood. 
It arises from nonlinear terms in the equations of motion, and from the lack of a sufficient number of constants of motion compared to the number of degrees of freedom of the system \cite{gutzwiller2013}. 
The hallmark of chaotic dynamics is the extreme sensitivity to initial conditions, whereby the trajectory of a system prepared in two infinitesimally different states evolves along two trajectories that diverge exponentially. 
Chaos plays a fundamental role in, for example, establishing the validity of classical statistical mechanics and thermodynamics, and a practical role in a wide range of applications, from weather forecasting to the design of tokamaks for nuclear fusion.

The usual description of quantum systems, in terms of states vectors that evolve according the Schr\"odinger equation, can appear puzzling when examined in the context of the chaotic behavior of the equivalent classical Hamiltonian. 
Consider for example two slightly different quantum states at time $t=0$, $|\psi_1(0)\rangle$ and $|\psi_2(0)\rangle$ having an initial  overlap $\Omega(0)=|\langle \psi_1(0)|\psi_2(0)\rangle|^2 = 1-\delta^2$, with $\delta \ll 1$. 
As time progresses, these states evolve according to the time evolution operation $U(t)$. 
The overlap at later times is thus $\Omega(t)=|\langle \psi_1(0)|U^{\dagger}(t)U(t)|\psi_2(0)\rangle|^2$. Since the time evolution is unitary, $U^{\dagger}(t)=U^{-1}(t)$, we find that $\Omega(t)=\Omega(0)=1-\delta^2$; i.e., the overlap remains constant at all times. 
Since the exponential divergence of trajectories typical of classical systems appears ruled out, does this mean that there cannot be chaos in quantum dynamics?

A more appropriate and illuminating comparison between classical and quantum dynamics is obtained by describing the classical system in terms of a density $f$ in phase space, and calculating its time evolution using the Liouville equation $i\frac{\partial f(t)}{\partial t}=\mathcal{L}f(t)$, where $\mathcal{L}$ is the Liouville operator \cite{peres1996}. 
One then finds that, given two initially overlapping densities $f_1(0)$ and $f_2(0)$, the Liouville equation for a conservative Hamiltonian system ensures that their overlap remains constant at all times \cite{koopman1931}. 
This property mirrors the quantum behavior described earlier, so now the question may be reversed: in what way, if at all, does classical chaos differ from the dynamics of quantum  systems?

The answer to this question can be rather subtle. 
At its heart, quantum mechanics requires that the classical phase space is coarse-grained into volumes of size $\hbar^N$ (with $N$ the number of degrees of freedom), and forbids specifying the state of the system to a precision finer than that. 
An illuminating example of how this affects the dynamics of chaotic systems was provided by Korsch and Berry \cite{korsch1981}, who analyzed a classically chaotic iterative map while varying the value of $\hbar$. 
In the classical limit ($\hbar \rightarrow 0$), the map `shreds' the initially smooth distribution into thin chaotic-looking `tendrils'. 
Conversely, when $\hbar$ becomes sizable on the scale of the map's effect, one finds that the distribution remains smooth and seems to lose its chaotic features, displaying instead a slow spread from the initial shape. 
More generally, since bounded quantum systems have a discrete spectrum, their dynamics exhibits quasiperiodic features that are at odds with the `true chaos' seen in classical systems \cite{casati1995}. 
Therefore, it is often stated that `quantum chaos' constitutes a new and unique type of dynamics \cite{berry1987}. 
Finally, quantum systems allow for interference effects that result in peculiar dynamical features, such as the development of structures in phase space at a scale smaller than the Planck constant, which are reached most quickly when the classical system is chaotic \cite{zurek2001}. 

The issue of how to observe and interpret signatures of chaos in quantum mechanics has profound repercussions on many important topics in physics.
For example, classical chaos underpins the ergodic hypothesis in statistical mechanics, and it is expected that its quantum equivalent plays a fundamental role in the thermalization of isolated quantum systems \cite{Deutsch1991, srednicki1994,rigol2008,eisert2015}. 
Chaos is also thought to be related to the issue of decoherence \cite{zurek1994}, which is crucial in the modern topic of quantum information science. 
There, one must answer the delicate question of whether an onset of chaos may harm the operation of a large-scale quantum computer \cite{georgeot2000,silvestrov2001,braun2002}. 
On the other hand, it has been suggested that the inherent ability of chaotic systems to quickly explore a vast configuration space can be used for the purpose of demonstrating `quantum supremacy' in multiqubit devices without error correction \cite{boixo2016}. 

\subsection{Experimental tests of quantum chaos}

Despite its broad and deep importance, experimental progress in `quantum chaos' is rare. 
Early work focused on the study of static and statistical properties of chaotic systems \cite{HaqPRL1982NED,LeviandierPRL1986,held1998observation}, such as the energy spectra of chaotic billiards implemented in semiconductor quantum dots \cite{marcus1992,wilkinson1996,folk1996}. 
Even more rare is the ability to experimentally observe \emph{dynamical} chaos, i.e. signatures of chaos in the time evolution of quantum systems. 
Crudely speaking, this is because most quantum systems decohere and randomize for trivial reasons (noise, uncontrolled environments, etc.) over time scales that are too short for signatures of chaotic behavior to reveal themselves. 
Conversely, systems with long coherence times, such as ensembles of nuclear spins in liquids, can show signatures of chaos in the dynamics of their macroscopic magnetization \cite{lin2000}. 
The experimental state of the art for truly quantum chaotic dynamics is found in ensembles of cold gases \cite{hensinger2001, steck2001, chaudhury2009, manai2015}, whereas only very recently an experiment on three superconducting qubits has provided experimental insight into the link between chaos and thermalization in a small-scale quantum system \cite{neill2016}. 

What is still missing is an experimental study of the quantum signatures of chaos in an \emph{individual} quantum system. 
Such a study will be an important complement and extension to experiments conducted on ensembles of particles, since an individual quantum system allows a much broader choice of measurement strategies.
Although the chaotic dynamics we will describe in this paper are the result of the Hamiltonian evolution alone, the use of an individual quantum  system will allow us in the future to explore the interplay between the emergence of chaos and the measurements performed on the system. 
Theoretical studies \cite{sanders1989,eastman2016} predict that the measurement strength can be used as an additional experimental knob to tune the chaoticity of the system's dynamics.
The measurement strength on a single object can be tuned continuously \cite{katz2006,hatridge2013} from projective single-shot readout \cite{blatt1988,elzerman2004,vijay2011} to arbitrarily weak measurements, partial wave function collapse \cite{guerlin2007} and even measurement reversal \cite{katz2008}. 
Strings of individual measurement outcomes could be analyzed with sophisticated statistical techniques to extract the most accurate information on the trajectory of the quantum  object \cite{gammelmark2013}, providing unprecedented insights into the chaotic dynamics of a monitored quantum system. 
Variable-strength measurements have already been experimentally demonstrated in the $^{31}$P donor system \cite{muhonen2017}.

Moreover, in the realm of classical computation, it has been recently shown that a network of \emph{individually} chaotic electronic components can solve computationally hard problems faster than one using nonchaotic elements \cite{kumar2017}. 
The construction of a single quantum chaotic system amenable to networking and controlled interactions could provide insights into how the equivalent quantum circuit would perform in complex computational problems.  

\begin{figure}
\includegraphics[width=\columnwidth]{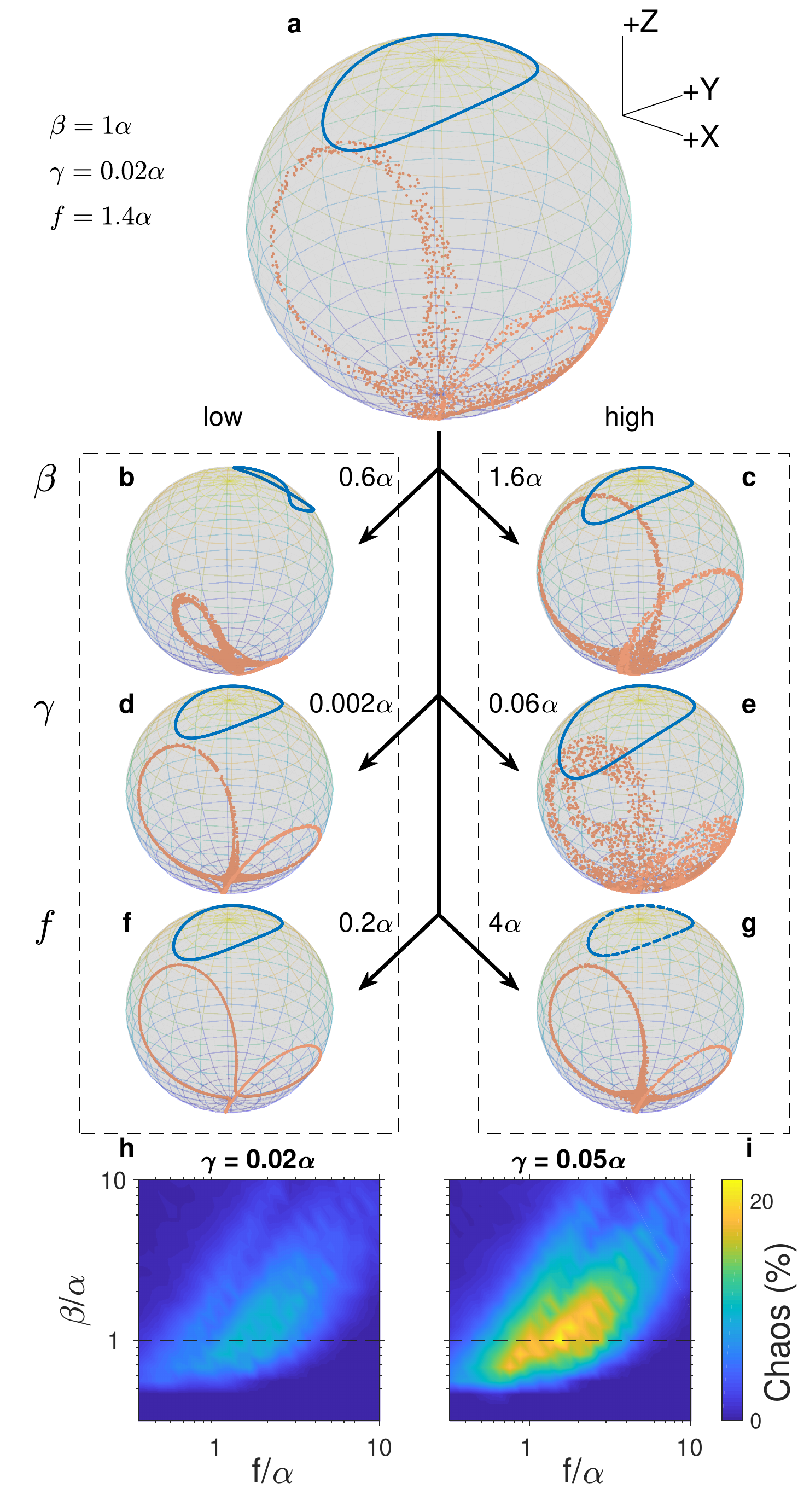}
\caption{Chaotic dynamics of the classical driven top [Eq.~\eqref{eq: H_top_classical}]. 
(a) Stroboscopic map of two trajectories of $\mathbf{L}$ corresponding to initial conditions in a regular [blue (dark gray)] and chaotic [orange (light gray)] region, with parameters shown to the left.
(b)-(g) Same as (a), but where a single parameter is varied.
(b),(c) Modifying the quadratic interaction strength $\beta$ shrinks the chaotic region and displaces the enclosed regular regions.
(d),(e) Increasing (decreasing) the periodic drive strength $\gamma$ leads to an enlarged (reduced) chaotic region.
(f),(g) Shifting the drive frequency $f$ away from precession frequencies at the boundaries of regular regions results in a reduced chaotic region.
(h),(i) Chaotic percentage of total phase space as a function of quadratic interaction strength and periodic drive frequency for weak [(h), $\gamma = 0.02 \alpha$] and strong [(i), $\gamma = 0.05 \alpha$] periodic drive strength (see Appendix~\ref{app: methods} for details of the calculation). Dashed horizontal lines indicate $\beta = \alpha$. 
}
\label{fig: Figure 1}
\end{figure}
\subsection{Quantum chaotic driven top} 

Here we present a detailed and quantitative proposal to experimentally realize a single-atom version of one of the best studied quantum chaotic systems, the `kicked top' \cite{haake2000}. 
For experimental convenience, we will focus on the case where the top is periodically driven, instead of kicked with $\delta$ functions. 
This system becomes chaotic in the presence of a term in the Hamiltonian that is quadratic in the angular momentum, and has a classical Hamiltonian of the form
\begin{equation} \label{eq: H_top_classical}
\mathcal{H}_\mathrm{classical} = \alpha L_\mathrm{z}+\beta L_\mathrm{x}^2+ \gamma \cos \left(2 \pi f t \right) L_\mathrm{y},
\end{equation}
with angular momentum $\mathbf{L}=\left(L_\mathrm{x} \hspace{0.5em} L_\mathrm{y} \hspace{0.5em} L_\mathrm{z}\right)^\mathrm{T}$ ($|\mathbf{L}|=$ constant), $\alpha$, $\beta$, and $\gamma$ proportionality constants, and $f$ the frequency of the drive.
The size and shape of the regular and chaotic regions in classical phase space are determined by the Hamiltonian parameters  $\alpha$, $\beta$, $\gamma$, and $f$  (Fig.~\ref{fig: Figure 1}). 
A sizable region of chaos is found with linear and quadratic interactions of similar strength ($\beta \approx \alpha$), a sufficiently strong periodic drive ($\gamma \gtrsim 0.02 \alpha$), and a drive frequency close to the range of  resonance frequencies of the nonlinear system ($f \approx 1.4 \alpha$).
These are the conditions we seek to reproduce in the quantum driven top.

The obvious quantum equivalent of a classical spinning top is a spin. 
The challenge here is to find a spin system whose Hamiltonian maps onto that of the chaotic driven top.
This requires in particular a quadratic term in the Hamiltonian, which is only possible for a spin quantum number $I>1/2$. 
Moreover, a larger spin is crucial in comparing its dynamics to the structure of the classical phase space, as its smaller \textit{relative} uncertainty spread allows for better localization of the quantum state in a certain area of interest in classical phase space (see Appendix~\ref{app: spin_properties} for further details). 
Using a single spin with access to high-fidelity single-shot state readout and/or variable strength weak measurements opens up the largely unexplored area of the interplay between chaos and quantum measurements. 
Lastly, it is of paramount importance that this system does not lose coherence for trivial reasons, unrelated to chaos, on time scales short compared to the chaotic dynamics. 
This requires a long intrinsic quantum coherence time of the system. 

\subsection{Experimental platform}

Our proposed system meeting these requirements is the nuclear spin of a heavy group-V substitutional donor in isotopically enriched $^{28}$Si \cite{itoh2014}. 
The lightest group-V donor in silicon, $^{31}$P, has been extensively studied in the context of quantum information processing \cite{kane1998}, since it naturally contains two quantum bits, the electron (with spin $S = 1/2$) and the $^{31}$P nucleus (with spin $I = 1/2$). 
High-fidelity single-shot readout \cite{morello2010single,pla2013high}, coherent operation \cite{pla2012single,pla2013high}, mutual entanglement \cite{dehollain2016} and variable-strength measurements \cite{muhonen2017} have been experimentally demonstrated. 
When implanted \cite{donkelaar2015} in isotopically enriched $^{28}$Si, these single-atom spins exhibit outstanding coherence times \cite{muhonen2014storing} (up to 35 seconds for the nuclear spin) and control fidelities \cite{muhonen2015,dehollain2016optimization}. 
This suggests that donor spin systems would be ideal platforms to study the subtle effects of dynamical chaos and its interplay with quantum measurement, if it were possible to engineer a suitable spin Hamiltonian. 
This is not the case with $^{31}$P, since its spin value of 1/2 forbids the presence of quadratic terms in the Hamiltonian. 
Heavier donors, such as $^{75}$As, $^{121}$Sb, $^{123}$Sb and $^{209}$Bi, all have nuclear spins $I > 1/2$, which allows the existence of a nuclear quadrupole interaction, scaling quadratically in the spin operators. 
Below we show that, under realistic conditions of quadrupole interaction and periodic drive, a heavy group-V donor can become a single-atom solid-state implementation of a chaotic driven top.

\section{Results} \label{sec: results}

\begin{figure*}
\includegraphics[width=\textwidth]{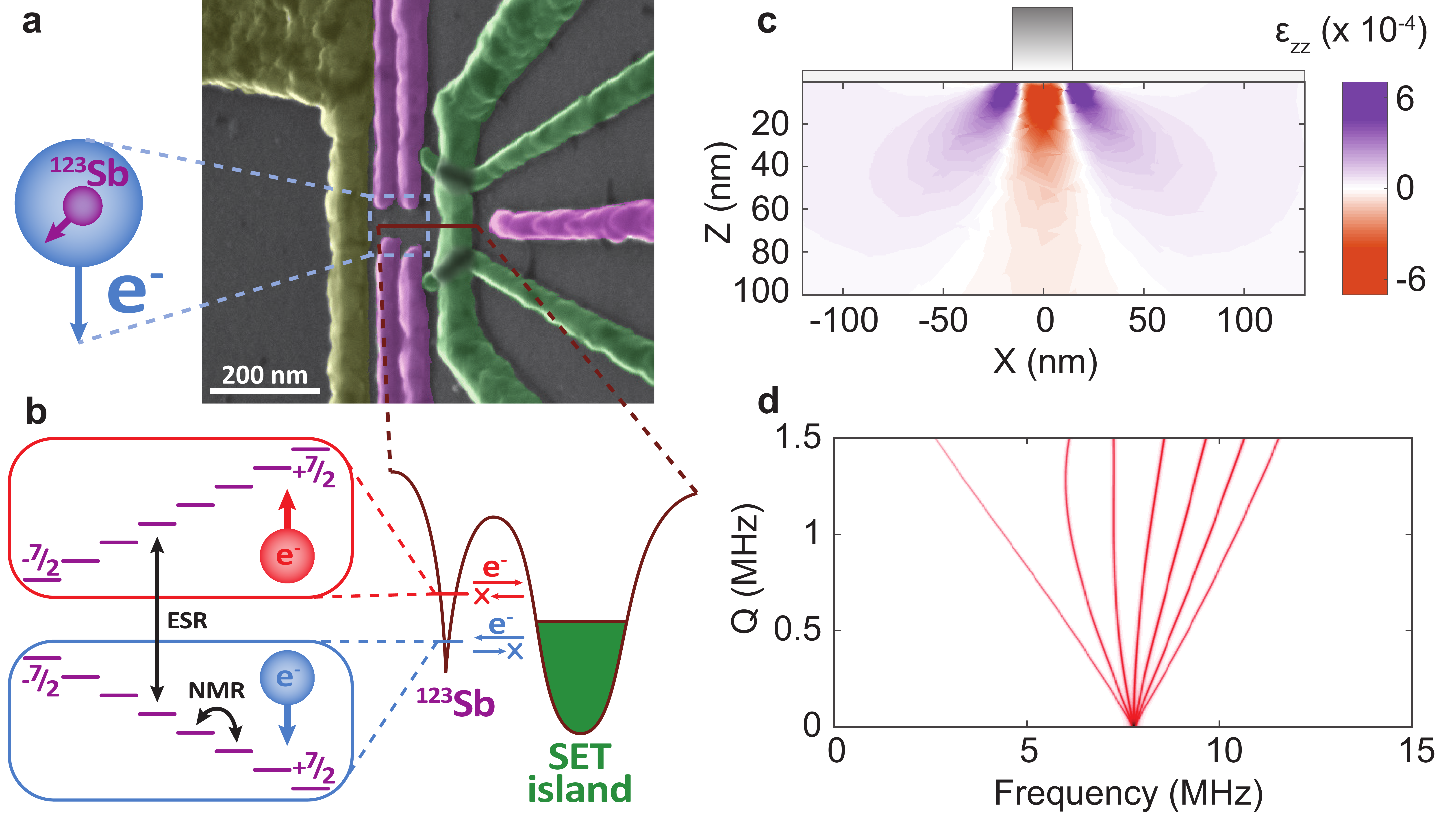}
\caption{Experimental implementation of the quantum driven top. 
(a) Scanning electron microscope image of a typical device. 
\Sb donors can be implanted in the \Sipure epilayer at the indicated position. Their electrochemical potential is controlled by electrostatic gates (false colored in purple), and a single-electron transistor (green) is fabricated in the vicinity of the donor implant area, to provide time-resolved electron spin readout via spin-dependent tunneling. A broadband microwave antenna (yellow) provides oscillating magnetic fields to excite the electron (ESR) and nuclear (NMR) resonances, and to periodically drive the nuclear spin.
(b) The 16 energy levels of the \Sb donor (spacings not to scale), separated into the electron spin-down $\ket{\downarrow}$ (blue) and spin-up $\ket{\uparrow}$ (red) manifold. For electron spin readout, the donor electrochemical potential is tuned such that only the $\ket{\uparrow}$ state can tunnel out of the donor, while only the $\ket{\downarrow}$ state can tunnel back onto it. The SET is biased such that the current is nonzero when the donor is ionized. (c) Finite-elements model of the strain induced in silicon at low temperatures by a $30 \times \SI{30}{\nano \meter \squared}$ aluminum gate, placed on top of a 5~nm thick SiO$_2$ dielectric. The maximum strain approaches 0.1\%, a value sufficient to generate strong quadrupole interaction enabling implementation of the quantum driven top. (d) NMR spectrum versus quadrupole interaction strength $Q$. Donor is ionized $\left(A=0\right)$, $B_0 = \SI{1.4}{T}$ and orthogonal to the direction of quadrupole interaction, $\eta = 0$. A non-zero quadrupole interaction leads to an unique spectroscopic fingerprint. The long expected lifetime of the nuclear spin states allows for very precise measurement of the spectral lines, which in turn enables accurate determination of the quadrupole interaction; see also Appendix~\ref{app: spectroscopy}.}
\label{fig: Figure 2}
\end{figure*}

\subsection{A large nuclear spin donor as quantum driven top} \label{subsec: }

The spin Hamiltonian of group-V donors in silicon, in the presence of a static magnetic field $B_0$ in the $z$ direction and an oscillating magnetic field $B_1$ at frequency $f$ in the $y$ direction, reads 
\begin{equation} \label{eq: H_donor}
\begin{aligned}
\mathcal{H} = &\left(\gamma_\mathrm{e}S_\mathrm{z}-\gamma_\mathrm{n}I_\mathrm{z}\right)B_0+A \hspace{0.25em} \mathbf{S}\cdot \mathbf{I} + \mathcal{H}_\mathrm{Q}\\
 &+ \left(\gamma_\mathrm{e}S_\mathrm{y}-\gamma_\mathrm{n}I_\mathrm{y}\right)B_1 \cos \left( 2 \pi f t\right) , 
\end{aligned}
\end{equation}
where $\gamma_\mathrm{e}$ and $\gamma_\mathrm{n}$ are the electron and nuclear gyromagnetic ratios (their magnetic moments have opposite sign), the electron spin $S=1/2$ is described by the vector of operators $\mathbf{S}=\left(S_\mathrm{x} \hspace{0.5em} S_\mathrm{y} \hspace{0.5em} S_\mathrm{z}\right)^\mathrm{T}$, the nuclear spin $I$ is described by $\mathbf{I}=\left(I_\mathrm{x} \hspace{0.5em} I_\mathrm{y} \hspace{0.5em} I_\mathrm{z}\right)^\mathrm{T}$, $A$ is the hyperfine interaction between electron and nuclear spin (assumed to be isotropic), and $\mathcal{H}_\mathrm{Q}$ accounts for the nuclear quadrupole interaction (discussed below).

Experimentally implemented via a broadband on-chip antenna, the oscillating $B_1$ field allows for coherent control of the spins through electron spin resonance (ESR) and nuclear magnetic resonance (NMR). 
This enables the application of numerous techniques for tomography and characterization of the spin system \cite{wolfowicz2016}. 

The hyperfine interaction $A$ couples the electron and the nuclear spins, and can be approximated by an effective interaction $A S_\mathrm{z} I_\mathrm{z}$ in the device operating regime where $\gamma_\mathrm{e}B_0 \gg A$.
This introduces a dependence of the ESR frequency on the state of the nucleus.
In turn, this allows the measurement (and consequent initialization by measurement) of the nuclear spin state by observing at what frequency the electron spin responds to a resonant microwave excitation using electron spin read-out via a standard spin-to-charge conversion technique~\cite{pla2013high} (Fig.~\ref{fig: Figure 2}).

The nuclear quadrupole moment is caused by the non-spherical charge distribution of the nucleus. 
This quadrupole interacts with electric field gradients to introduce a new term in the spin Hamiltonian~\cite{KauRMP79}. 
In general, the tensor describing the electric field gradient can be diagonalized to $\diag\hspace{-.2em}\left(V_\mathrm{x'x'}, V_\mathrm{y'y'}, V_\mathrm{z'z'}\right)$ by an appropriate choice of coordinate frame $(x',y',z')$, where $V_\mathrm{i j}=\tfrac{\partial^2 V}{\partial {i} \partial {j}}$, $\left(i,j \in x',y',z'\right)$ are the partial second derivatives of the electrostatic potential $V$, and $|V_\mathrm{x'x'}|\leq|V_\mathrm{y'y'}|\leq|V_\mathrm{z'z'}|$.
The expression for the nuclear quadrupole interaction is then simplified to~\cite{SliPMR90}
\begin{equation} \label{eq: H_Q}
\mathcal{H}_\mathrm{Q} = Q\left(I_\mathrm{z'}^2-\frac{I^2}{3}+\frac{\eta}{3}\left(I_\mathrm{x'}^2-I_\mathrm{y'}^2\right)\right),
\end{equation}
where $Q$ is the effective quadrupole interaction strength, which scales linearly with both $V_\mathrm{z'z'}$ and the nuclear quadrupole moment $Q_\mathrm{n}$, and the asymmetry parameter $\eta = \frac{V_\mathrm{x'x'}-V_\mathrm{y'y'}}{V_\mathrm{z'z'}}$ quantifies the deviation from axial symmetry of the electric field gradient ($0 \leq \eta \leq 1$, $\eta = 0$ corresponds to axial symmetry; see Appendix~\ref{app: quadrupole} for further details on quadrupole interaction).

The important features of the quadrupole interaction are that it is quadratic in the spin operators and has a preferred quantization axis. It is often the case that the electric field gradient tensor has approximately axial symmetry ($\eta = 0$) \cite{SliPMR90}. If, in addition, the static magnetic field $B_0$ can be oriented in an arbitrary direction (for example using a 3-axis vector magnet), the linear and quadratic terms in the spin Hamiltonian can be made orthogonal.
Ignoring the static energy offset $QI^2/3$, and assuming for simplicity that the symmetry axis of the electric field gradient is orthogonal to the periodic driving field $B_1$, the Hamiltonian of a donor with nuclear spin $I > 1/2$ takes the form 
\begin{equation} \label{eq: H_top_quantum}
\mathcal{H}_\mathrm{quantum} = \left(\gamma_\mathrm{n}B_0 \pm \tfrac{1}{2}A \right) I_\mathrm{z} + Q I_\mathrm{x}^2 + \gamma_\mathrm{n} B_0 \cos \left( 2 \pi f t \right) I_\mathrm{y}.
\end{equation}  
Therefore, $\mathcal{H}_\mathrm{quantum}$ represents the quantum equivalent of the Hamiltonian of a classical periodically driven top [Eq.~\eqref{eq: H_top_classical}].

\subsection{Donor parameters and preferred operation regimes} \label{sec: donor_parameters}

\subsubsection{Realising a quantum driven top in the laboratory frame}

We now estimate the parameters of the spin Hamiltonian of the quantum driven top, in order to compare them to the parameters that are known to lead to chaotic dynamics in the equivalent classical case.  

The key parameter values of the group-V donors are summarized in Table~\ref{tab: donor_parameters}. 
The value of $I$ increases with atomic mass, while the hyperfine interaction $A$ has a non-monotonic behavior, with a significant jump for the heavy $^{209}$Bi donor. 
Large $I$ values are desirable to reduce the relative quantum uncertainty of the spin state (see Appendix~\ref{app: spin_properties} for details), whereas from the analysis of the classical driven top we know that interesting chaotic dynamics arises when linear and quadratic terms in the spin Hamiltonian are of comparable strength.

For operation in the neutral charge state, the linear term has strength $\gamma_n B_0 \pm A/2$, which becomes very large in the case of \Bi where $A=\SI{1475.4}{\mega \hertz}$. 
The $\pm A/2$ contribution can be removed by operating in the ionized state~\cite{pla2013high}.
The nuclear Zeeman term is minimized by operating at low $B_0$, with the caveat that reducing $B_0$ affects the electron readout and initialization fidelity. 
Using a minimum attainable value of $B_0 = 0.5$~T and ionized donors, the linear term in the spin Hamiltonian thus takes values of order 3~MHz.

Next, we wish to obtain a comparable value for the quadratic term, which is achieved by maximizing the strength of the quadrupole interaction, and hence the electric field gradient [Eq.~\eqref{eq: H_Q}]. 
Recent experiments~\cite{ThoAIPA15,park2016,PlaarXiv16} found that aluminum gates on top of a Si/SiO$_2$ stack cause strain in the silicon; finite-element models indicate that strain values approaching 0.1\% can be expected [Fig.~\ref{fig: Figure 2} (c)].
This strain-induced disruption of the cubic lattice symmetry under the metal results in an electric field gradient at the donor site. 
Such an electric field gradient is further enhanced at the nucleus due to rearrangement of the closed electronic shells of the donor (the Sternheimer anti-shielding effect~\cite{KauRMP79,FeiPR69}). 
These effects are present in both the neutral and the ionized charge state, albeit with different strengths.

Measurements of the quadrupole interaction strength of group-V donors in strained silicon, especially near a nanostructure interface, are limited and have only been conducted very recently~\cite{FraPRL15, FraPRB16, MorIOPN16}.
For ionized group-V donors, the available data only allow an order-of-magnitude estimate of $Q$ in the hundreds of kHz range (see Appendix~\ref{app: quadrupole} for an extensive review). 

\begin{table}
\caption{\label{tab: donor_parameters} Parameters of group-V donors in silicon. 
Ground $1s$ state binding energies taken from Ref.~\onlinecite{ramdas1981spectroscopy}. 
Hyperfine interaction $A$ taken from Ref.~\onlinecite{feher1959electron}. 
Nuclear gyromagnetic ratio $\gamma_\mathrm{n} = \mu \mu_\mathrm{n}/I$ is calculated from the nuclear magnetic moment $\mu$ given in units of nuclear magneton $\mu_\mathrm{n} =$ \SI[per-mode=symbol]
{7.62}{\mega \hertz \per \tesla} in Ref.~\onlinecite{stone2014table}.
Minimum and maximum values of nuclear quadrupole moment $Q_\mathrm{n}$ are given, based on the range of values reported in Ref.~\onlinecite{stone2014table}.}
\begin{ruledtabular}
\begin{tabular}{cccccc}
donor & $I$ & 1s binding energy & $A$ & $\gamma_\mathrm{n}$ & $Q_\mathrm{n}$ \\
& & ($\si{\milli \electronvolt}$) & ($\si{\mega \hertz}$) & ($\si[per-mode=symbol]{\mega \hertz \per \tesla}$) & $(10^{-28} \si{\square\meter})$ \\
\phosphor & $1/2$ & 45.59 & 117.53 & 17.26 & - \\
\As & $3/2$ & 53.76 & 198.35 & 7.31 & 0.314\\
\Sblight & $5/2$ & 42.74 & 186.80 & 10.26 & [-0.36 , -0.54]\\
\Sb & $7/2$ & 42.74 & 101.52 & 5.55 & [-0.49 , -0.69]\\
\Bi & $9/2$ & 70.98 &  1475.4 & 6.96 & [-0.37 , -0.77]\\
\end{tabular}
\end{ruledtabular}
\end{table}

Upon comparing the classical and quantum systems, the coefficient $\beta$ in the classical Hamiltonian must be compared to $Q I$ in the quantum case (Appendix~\ref{app: compare_H}). 
This makes the hundreds of kHz range for $Q$ rather promising, since for large $I$ the corresponding $Q I \sim \SI{1}{\mega \hertz}$ is within an order of magnitude of the linear interaction strength ($\gamma_\mathrm{n}B_0 \sim \SI{3}{\mega \hertz}$).
If this proves to be insufficient, strain engineering could be deployed to further increase the electric field gradient to reach the target value of $Q I \sim \SI{3}{\mega \hertz}$. This can be achieved either in MOSFET structures \cite{thompson2006} or in Si/SiGe devices, where the ability to electrically detect a single dopant atom coupled to a quantum dot has also been recently demonstrated \cite{foote2015}. 
All the above options will deliver a fixed value of strain, set by the thermal expansion of the metallic electrodes and/or the built-in strain in the substrate. 
As a next step in experimental sophistication and control, one could consider fabricating on-chip piezoactuators to dynamically control the strain \cite{dreher2011}, allowing the study of the chaotic dynamics of a nuclear spin as a function of its Hamiltonian parameters. 
Finally, another option to tune \textit{in} situ the quadrupole splitting could be to distort the electron wave function using strong voltages on gates placed above the donor, to the point where the electron wave function is significantly displaced from the nuclear site, potentially generating a substantial electric field gradient. 
This type of electron wave function distortion has been discussed in numerous papers \cite{calderon2006,lansbergen2008,tosi2017}, but no calculation of the resulting nuclear quadrupole splitting in the case of a $I>1/2$ nucleus has been performed to date. 

The broadband antenna near the donor can be used to apply a radio-frequency periodic drive. 
Previous work has been conducted with drive strengths up to $B_1 \sim \SI{2}{\milli \tesla}$~\cite{pla2013high}, which correspond to radio-frequency powers of order 0.5~mW (at the chip). 
Those values were sufficient to achieve high-fidelity coherent control of the \phosphor nuclear spin qubit. Here, we wish to compare $\gamma_\mathrm{n} B_1$ to the classical parameter $\gamma$. 
Chaos arises when $\gamma \approx 0.02 \alpha$ in the classical model. 
For $B_0 \sim 0.5$~T and thus $\alpha \sim 3$~MHz, this implies $B_1 \sim 10$~mT. 
Assuming the same setup and antenna as Ref.~\onlinecite{pla2013high}, this value would require $\sim 10$~mW radio-frequency power at the chip. 
This is a very high value for operation at millikelvin temperatures, but we note that the broadband microwave antenna is terminated by a short circuit, constituting (ideally) a fully reflective load. Therefore, only a small fraction of the incident power is actually dissipated on the chip, while the rest is reflected and dissipated at stages of the setup with large cooling powers.
Alternatively, high $B_1$ values with low incident power could be obtained by using $LC$ resonators.

\subsubsection{Realizing a quantum driven top in the rotating frame}

The Hamiltonian of the periodically driven top described so far was defined in the laboratory frame. 
An alternative approach is to describe it in the rotating frame, defined by the frequency of an oscillating field. 
This results in a system `dressed' by a continuous radio-frequency field, at a frequency that matches the nuclear Zeeman interaction strength ($f_\mathrm{RF}=\gamma_\mathrm{n}B_\mathrm{0}$).
This is a well-established method that originates from quantum optics \cite{mollow1969} and has recently been extended to microwave frequencies \cite{timoney2011}, including with the electron spin of the \phosphor donor \cite{laucht2017dressed}. 
Here we analyze its application to the higher-dimensional nuclear spin of the heavier group-V donors. We consider a Hamiltonian for the ionized nucleus $\left(A=0\right)$ of the form:

\begin{equation} \label{eq: H_top_quantum_AM_mod}
\begin{aligned}
& \mathcal{H}_\mathrm{quantum,RF} = \gamma_\mathrm{n}B_\mathrm{0} I_\mathrm{z}+Q I_\mathrm{x}^2 + \\
& \left[ \gamma_\mathrm{n}B_\mathrm{1,I} \cos{\left(2\pi f_\mathrm{RF}t\right)} 
+\gamma_\mathrm{n}B_\mathrm{1,Q}\cos{\left(2\pi f t \right)} \sin{\left(2\pi f_\mathrm{RF} t\right)}\right]I_\mathrm{y}.
\end{aligned}
\end{equation}

Here $B_\mathrm{1,I}$ and $B_\mathrm{1,Q}$ are the in-phase and quadrature amplitudes of an $\mathrm{IQ}$-modulated radio-frequency drive at frequency $f_\mathrm{RF}=\gamma_\mathrm{n}B_\mathrm{0}$, with an additional amplitude modulation applied to the in-phase component of the drive, at a frequency $f$.  
Switching the system to the rotating frame and applying the rotating wave approximation (RWA) will effectively remove the static Zeeman energy, while reintroducing a linear term which scales with the strength of the continuous drive at frequency $f_\mathrm{RF}=\gamma_\mathrm{n}B_0$ (see Appendix~\ref{app: RWA} for a derivation). 
Now the Hamiltonian reads
\begin{equation} \label{eq: H_quantum_top_RWA}
\begin{aligned}
\mathcal{H}_\mathrm{quantum, RWA} = & - \frac{1}{2} \gamma_\mathrm{n} B_\mathrm{1,I}  I_\mathrm{y}\\
& -\frac{1}{2}QI_\mathrm{z}^2+\frac{1}{2}\gamma_\mathrm{n} B_\mathrm{1,Q}\cos{\left(2\pi f t\right)}I_\mathrm{x},
\end{aligned}
\end{equation}

which is, up to a trivial rotation, equivalent to the quantum driven top [Eq.~\eqref{eq: H_top_quantum}].

Engineering a dressed system and considering this in the rotating frame has several important benefits for the experimental feasibility of our proposal. 
Firstly, the linear term in the new Hamiltonian, $\alpha = \gamma_\mathrm{n} B_\mathrm{1,I}/2$, is continuously (and rapidly, if desired) tunable all the way to zero, and up to a maximum set by the strongest attainable oscillating field strength ($B_\mathrm{1} \sim 10$ mT as assumed earlier), corresponding to $\alpha \sim \SI{30}{\kilo \hertz}$. 
This means that we can access a strongly chaotic regime, where the quadratic term $\beta = QI$ is comparable to the linear term, with quadrupole interaction strengths of only $Q \sim \SI{10}{\kilo \hertz}$.

Second, $\mathrm{IQ}$ modulation combined with amplitude modulation is a standard microwave control technique, which allows full and independent control over the strength of the periodic drive at frequency $f$ between 0 and the maximum strength of the linear interaction $\tfrac{1}{2}\gamma_\mathrm{n}B_\mathrm{1,I} = 0 \sim \SI{30}{\kilo \hertz}$. 
This opens up a new parameter regime of very strong periodic drive, with increased size of classical chaotic regions (see Appendix~\ref{app: RWA}), which would be challenging to obtain  in the laboratory frame.

Lastly, the constraint of orthogonality between axis of quadrupole interaction and direction of periodic drive, as imposed by the classical system [Eq.~\eqref{eq: H_top_classical}], is relaxed in the rotating frame under the RWA (see Appendix~\ref{app: RWA}). 
This is important, since now the axis of the static Zeeman field $B_0$ only needs to be perpendicular to the plane defined by the directions of quadrupole interaction and periodic drive, regardless of the relative angle between the latter two. This is easily achievable using a 3D vector magnet.

Overall, moving to the rotating frame can allow exploring a wider parameter space, but the actual timescale of dynamical phenomena will be scaled down by a factor $\sim 100$ (for $\tfrac{1} {2}\gamma_\mathrm{n}B_\mathrm{1,I} \sim \SI{30}{\kilo \hertz}$ vs $\gamma_\mathrm{n}B_\mathrm{0} = \SI{2.8}{\mega \hertz}$ at $B_\mathrm{0}=\SI{0.5}{T}$). For example, the period of dynamical tunneling [Fig.~\ref{fig: figure 4}(c) for the case of the laboratory frame] will become $\sim 100~\mu$s for the same choice of relative parameter strengths. This remains several orders of magnitude faster than the expected intrinsic coherence time of the nuclear spin, noting also that the technique of dressing a spin with a driving field often yields an extra order of magnitude in coherence time \cite{laucht2017dressed}.

\subsubsection{Summary}

In summary, these estimates suggest that the \Sb and \Bi donors in silicon are suitable candidates to implement the quantum driven top, due to their high spin quantum number ($I=7/2$ and $I=9/2$, respectively), low nuclear gyromagnetic ratio ($\gamma_\mathrm{n}=\SI{5.55}{\mega \hertz \per \tesla}$ and $\gamma_\mathrm{n}=\SI{6.96}{\mega \hertz \per \tesla}$), and, in the case of \Sb, low hyperfine coupling strength ($A=\SI{101.52}{\mega \hertz}$). 
\Sb has the additional advantage that its suitability for ion implantation is well documented \cite{schenkel2006}: after low-energy implantation and high-temperature rapid thermal anneal, the Sb atoms are fully activated, and the implantation damage to the silicon lattice is thoroughly repaired. Recent work \cite{weis2012} suggests that the implantation damage can be efficiently repaired also in the case of \Bi, although the electrical activation yield remains lower than that of Sb.
The attainable quadrupole interaction is not well known, but recent work \cite{FraPRL15,FraPRB16} 
indicates that it can plausibly reach a comparable value to the nuclear Zeeman term in low  ($B_0 \sim \SI{0.5}{\tesla}$) static magnetic field. 
Furthermore, dressing the system at the nuclear Zeeman frequency $\gamma_\mathrm{n}B_\mathrm{0}$ replaces the linear interaction strength by the drive strength. 
This lowers the linear interaction strength, and thereby the minimum quadrupole strength, by two orders of magnitude, which brings the parameter regime of similar strength linear and quadratic interactions within reach using current device technology only.
When combined with a strong oscillating magnetic field ($B_1 \sim \SI{10}{\milli \tesla}$), we conclude that the parameter range where the equivalent classical driven top behaves chaotic throughout sizable areas of its phase space is within reach. 
In what follows, we will concentrate our discussion on the use of \Sb as the model system to study quantum chaos in a single spin.

\subsection{Quantum versus classical dynamics: a comparison} \label{subsec: quantum_vs_classical}

\begin{figure}
\includegraphics[width=\linewidth]{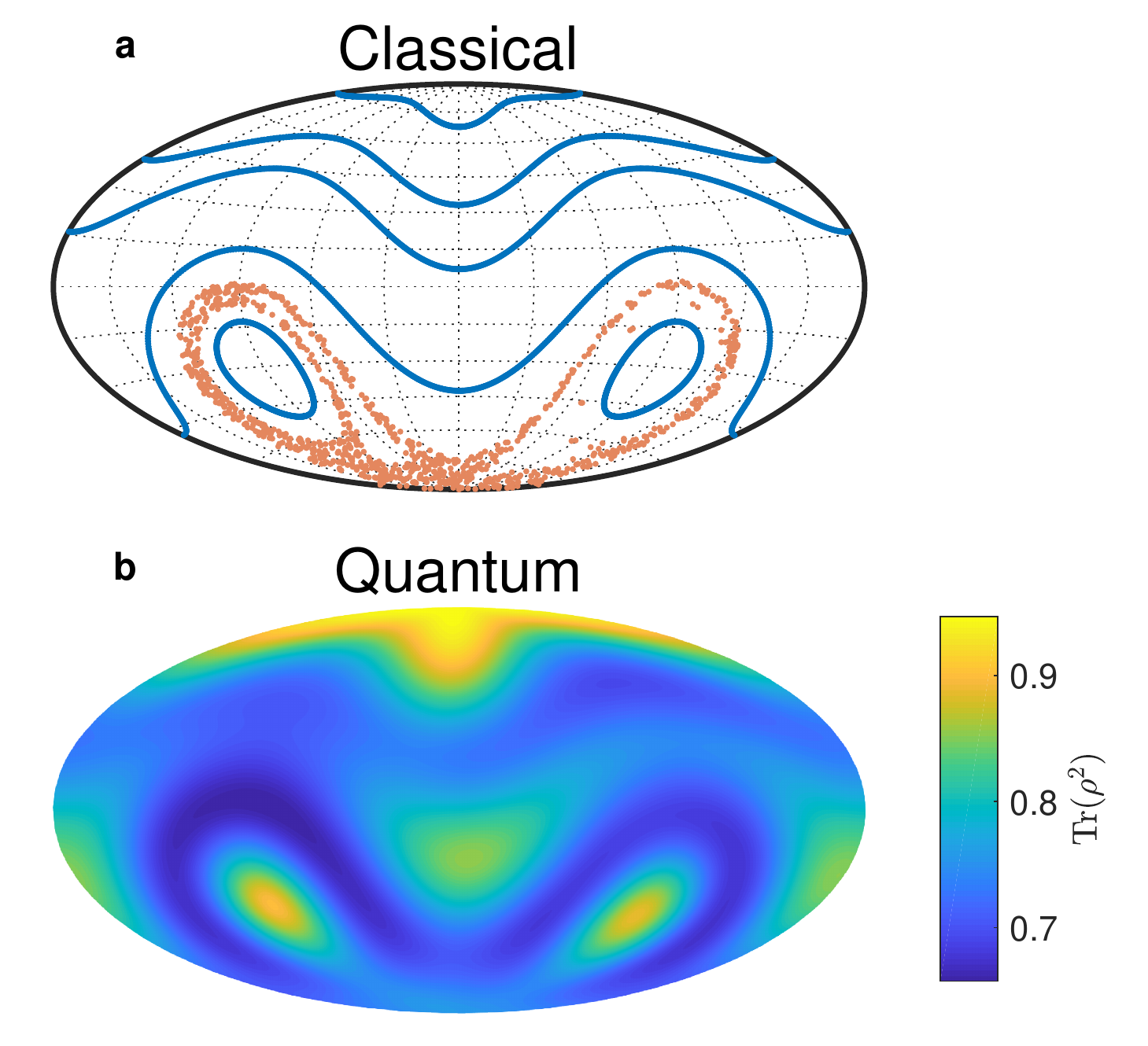}
\caption{
Comparison between classical Poincar\'e map and quantum state purity of a periodically driven top. 
The spherical surfaces of constant angular momentum are visualized by Hammer projections.
(a) Classical: Stroboscopic map of one chaotic trajectory [orange (light gray)] and six regular trajectories [blue (dark gray)] for $N=1000$ sampling periods, with the same simulation parameters as Fig.~\ref{fig: Figure 1}(a), i.e,. $\beta = \alpha$, $\gamma = 0.02\alpha$, $f=1.4\alpha$.
The chaotic region divides the three regular regions and contains regular trajectories within their respective regions.
(b) Quantum: Purity of spin coherent states $\ket{\theta, \phi}$ of an ionized \Sb atom with $I = 7/2$, $\gamma_n=\SI{5.55}{\mega \hertz \per \tesla}$, $B_0=\SI{0.5}{\tesla}$, $Q = 800 \pm 4$ $\mathrm{kHz}$, $B_1 = \SI{10}{\milli \tesla}$, $f = \SI{3.5}{\mega \hertz}$. 
The state purity is extracted from the density matrix $\rho$, which is obtained by evolving each spin coherent state for $N=1000$ drive periods while randomly perturbing $Q$ once per drive period, and averaging over 200 such evolutions (see Appendix~\ref{app: methods} for details of the calculation).
Similar behavior is observed upon varying $B_0$ or $B_1$ instead of $Q$ (Appendix~\ref{app: purity}).
The maps show a correspondence between the regular and chaotic regions of the classical phase space, and the purity of evolved spin coherent quantum states. 
}
\label{fig: figure 3}
\end{figure}

To illustrate the applicability of the \Sb system to the study of quantum chaos, we propose two types of experiments: one aimed at finding a correspondence between the classically chaotic driven top and its quantum counterpart, the other at demonstrating a violation of classical dynamics, exposing the true quantum nature of the system. 
In what follows, we focus on the system in the laboratory frame, as this puts the most stringent conditions on size and shape of the chaotic region in phase space of the classical equivalent system, however; the suggested experiments are equally well applicable to the system in the rotating frame.

\subsubsection{Decoherence as a precursor of chaos}

Classical chaos is characterized by an extreme sensitivity to perturbations, and as a consequence, neighboring trajectories with slightly different initial coordinates rapidly diverge. 
This is in stark contrast with quantum dynamics, where the discrete nature of the energy spectrum results in quasiperiodic behavior, leading to partial revivals of the initial quantum state instead (see Appendix~\ref{app: floquet} for details). 
However, quantum states evolving under slightly perturbed Hamiltonians do experience divergence. 
Since a quantum system is never truly isolated, interactions with its environment lead to unknown perturbations of the Hamiltonian, effectively entangling the system with its environment. 
The unknown nature of this process translates to the quantum state losing its purity, thus decohering into a mixed state.

Crucial to the quantum driven top, certain initial quantum states are more prone to decoherence, while others remain relatively unperturbed.
This behavior appears related to the high sensitivity of certain classical states to perturbations, but is caused here by a varying sensitivity of the time-evolution operator's eigenstates to perturbations (see Appendix~\ref{app: floquet} for details).
States containing more of these high-sensitivity eigenstates are therefore more susceptible to phase errors.
In line with this picture, Zurek and co-workers \cite{zurek1994} predict that the rate at which different initial quantum states decohere provides a mapping to the chaotic or nonchaotic nature of the corresponding classical system, with chaotic classical regions corresponding to more rapidly decohering initial quantum states.

\begin{figure*}[t]
\includegraphics[width=\textwidth]{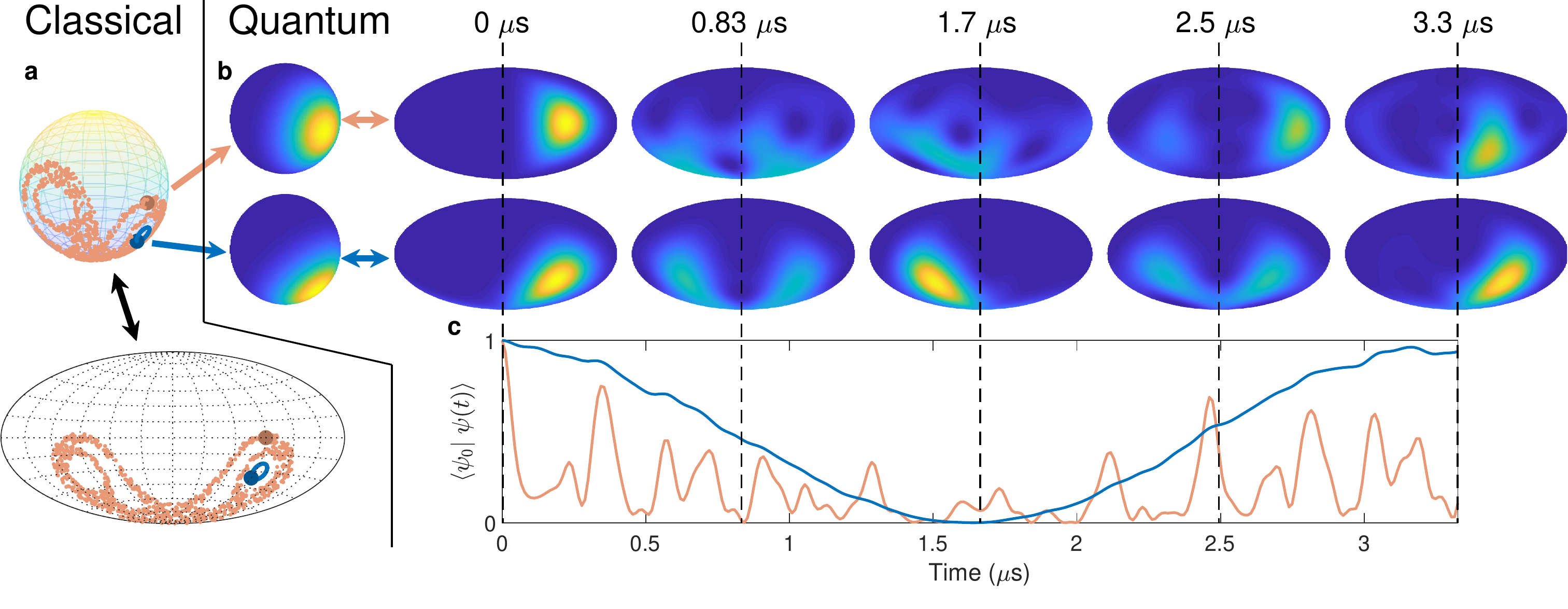}
\caption{Time evolution of the quantum driven top.
Two different initial spin coherent states are chosen in the directions of a classically chaotic region [orange (light gray)] and a classically regular region [blue (dark gray)]. 
Implementation in a $I = 7/2$ nuclear spin of an ionized \Sb donor in silicon with $\gamma_\mathrm{n} = \SI{5.55}{\mega \hertz \per \tesla}$, $B_0 = \SI{0.5}{\tesla}$, $Q = \SI{0.8}{\mega \hertz}$, $B_1 = \SI{10}{\milli \tesla}$, $f = \SI{5}{\mega \hertz}$. 
The equivalent implementation in the rotating frame, assuming $\gamma_n B_{1,\mathrm{I}}/2 ~\sim \SI{30}{\kilo \hertz}$ at a frequency $f_\mathrm{RF} = \gamma_\mathrm{n}B_0$, corresponds to a factor $\sim 100$ smaller values for $Q$ and $f$, i.e., $Q \sim \SI{8}{\kilo \hertz}$ and $f \sim \SI{50}{\kilo \hertz}$.
(a) Classical angular-momentum trajectories with initial states corresponding to orientations of spin coherent states and parameters matching the quantum simulations (Table~\ref{tab: classical_quantum_comparison}).
Trajectories are visualized by both a three-dimensional spherical plot (top) and Hammer projection (bottom), with enlarged dots representing the two initial angular-momentum coordinates.
(b) Husimi Q representation of the two spin coherent states at different moments in their evolutions (see Appendix~\ref{app: spin_properties} for details); movies of these time evolutions are part of the Supplemental Material \cite{supplemental}.
Color scale is constant across all panels, and varies between $0$ [dark blue (dark gray)] and $1/\pi$ [bright yellow (light gray)].
Top (bottom) row corresponds to spin coherent state oriented in the direction of the classically chaotic (regular) region in phase space, specified by orange (light gray [blue (dark gray)].
In both rows, the first color map is a three-dimensional view of the initial spin coherent state, equal to the first Hammer projection.
The subsequent Hammer projections show the evolution of the quantum state at later times.
The spin coherent state prepared in the classically chaotic region (top) displays a rapid dispersion over the phase space, while the classically regular spin coherent state (bottom) transfers back and forth between two classically regular regions.
This property is known as dynamical tunneling, and is in stark contrast with classical dynamics, where trajectories cannot cross closed regions.
(c) Overlap of the time-evolved state $\ket{\psi(t)}$ with its initial state $\ket{\psi(0)}$ for spin coherent states in classically chaotic [orange (light gray)] and regular [blue (dark gray)] regions.
Dynamical tunneling [blue (dark gray)] is revealed by a near-sinusoidal evolution of the overlap, returning to near unity.
This is in contrast to the evolution of the classically chaotic spin coherent state, where the overlap quickly decreases and shows no revival. 
Note that the rotating frame equivalent implementation will effectively multiply the time axis by a factor $\sim 100$ for the suggested parameters. 
As both in the laboratory and in the rotating frame implementation of the quantum driven top the dynamical tunneling time is multiple orders of magnitude smaller compared to the expected coherence time of order \SI{1}{\second}, no decoherence effects are included in this simulation.
}
\label{fig: figure 4}
\end{figure*}

The driven-top system can be used to verify this prediction, both through simulations and experiments. 
By evolving an initial state for a certain duration using the driven-top Hamiltonian [Eq.~\eqref{eq: H_top_quantum}], the resulting density matrix $\rho$ provides information about the degree of decoherence through its purity $\Tr(\rho^2)$. 
Decoherence is simulated by randomly varying a Hamiltonian parameter during the state's evolution, and calculating the state purity from the ensemble average of many final states, each obtained with a different randomized evolution (see App.~\ref{app: methods} for details). 
By sampling over all spin coherent states, a `purity map' of the quantum driven top is obtained, which we compare to its classical counterpart (Fig.~\ref{fig: figure 3}). 
The simulations highlight a correspondence between the classically-chaotic regions and quantum regions of strong decoherence, and between classically-regular regions and quantum regions of weak decoherence.
To experimentally verify these predictions, we aim to prepare spin coherent states (see App.~\ref{app: spin_properties} and App.~\ref{app: control} for details), evolve the system under the driven-top Hamiltonian, and finally reconstruct $\rho_\mathrm{final}$ using quantum state tomography.
Repeating this for different initial spin coherent states allows for experimental reconstruction of the `purity map', which can then be compared to the corresponding classical phase space. 

\subsubsection{Dynamical tunneling}

In the absence of a periodic drive, trajectories of the classical top are closed orbits confined to distinct regions in a two-dimensional phase space (the surface of a sphere due to $\left|\mathbf{L}\right|$ being a constant of motion).
Upon addition of a periodic drive, this behavior is largely kept intact, except near boundaries between the regular regions, where a chaotic behavior appears. 
The classically-separated regions of regular motion have an analogue in the corresponding quantum system, where they can be identified as regions of weak decoherence (Fig.~\ref{fig: figure 3}).
However, there is a fundamental difference between the quantum and the classical case. 
In the classical system, the Kolmogorov, Arnol'd and Moser (KAM) theorem ensures that a system initially prepared within one of the regular regions will remain on a periodic orbit within such a region. 
The quantum system, however, cannot be precisely localized within a certain region, due to the uncertainty principle. 
The `leakage' of the quantum wave function into a different regular region results in the phenomenon of dynamical tunneling, i.e. the tunneling of the quantum state between separate regular regions, in violation of the KAM theorem.

Dynamical tunneling manifests itself in the quantum driven top as a periodic oscillation of a spin coherent state between the two classically-regular regions associated with the quadratic interaction [Fig.~\ref{fig: figure 4}(a)]. 
In contrast, a spin coherent state prepared within a classically chaotic region rapidly spreads out and shows no apparent revival to a spin coherent state [Fig.~\ref{fig: figure 4}(b),~\ref{fig: figure 4}(c) and movies in Supplemental Material \cite{supplemental}]. 

Numerical simulations, conducted using Hamiltonian parameters appropriate for \Sb, clearly show the appearance of dynamical tunneling for spin coherent states prepared initially within the regions of classically regular periodic orbits [Fig.~\ref{fig: figure 4}(b),~\ref{fig: figure 4}(c)]. 
The predicted period of dynamical tunneling is $\sim \SI{3}{\micro \second}$ (see Appendix~\ref{app: dynamical tunneling} for dependence of tunnel rate on system parameters); this will increase by a factor $\sim 100$ to $\sim \SI{300}{\micro \s}$ upon considering the system in the rotating frame (assuming $\gamma_n B_{1,\mathrm{I}}/2 ~\sim \SI{30}{\kilo \hertz}$ and $Q$ is reduced by a factor 100 to $\sim \SI{8}{\kilo \hertz}$). 
This is a crucial result, since this period is orders of magnitude shorter than the dephasing time of a nuclear spin in \Sipure \cite{muhonen2014storing}, ensuring that the coherent dynamical tunneling oscillations can be observed in an experiment over unprecedented timescales.

\section{Conclusion and outlook} \label{sec: outlook}

In this paper, we have quantitatively described a proposal for the realization of a \emph{single} quantum chaotic system, based upon the nuclear spin of a substitutional group-V donor in silicon. 
In particular, we have shown that, with realistically achievable parameters, the $I=7/2$ nucleus of a $^{123}$Sb donor can exhibit the whole spectrum of features of interest in the study of experimental quantum chaos, from state-dependent decoherence to dynamical tunneling. 
The experimental verification of these predictions would constitute the first observation of quantum chaos in an individual physical system. Such an achievement would reinvigorate the fundamental study of quantum-classical boundaries by providing a well-defined and exquisitely controllable experimental test bed. 

In terms of applications, one could envisage laying out and operating individually chaotic $^{123}$Sb nuclei in the same types of multispin architectures that are being extensively studied in the context of quantum information processing with $^{31}$P donors \cite{hill2015,ogorman2016,pica2016,tosi2017}. 
Substituting a simple $I=1/2$ spin with a multilevel system like $^{123}$Sb could allow the study of quantum information processing where the information is encoded in an intrinsically chaotic system. 
This would be a different and complementary approach to the one taken, e.g., in superconducting systems, where it is the nature of the interaction between multiple qubits that produces a chaotic dynamics \cite{neill2016,boixo2016}. 
Rather, it could constitute the quantum version of a type of analog computation that has started to show promise in the context of classical neural networks, where having individually chaotic elements can speed up the solution of complex problems \cite{kumar2017}. 

\vspace{\baselineskip}

The supporting data for this work are available at Research Data Australia \cite{dataset}.

\section*{Acknowledgments}
We thank A. Laucht, F.A. Mohiyaddin, V. Schmitt, and G. Tosi for suggestions and comments. 
This work was funded by the Australian Research Council Discovery Projects No. DP150101863 and No. DP180100969. 
V.M. acknowledges support from a Niels Stensen Fellowship.
V.M. and S.A. performed the numerical simulations. H.F. wrote the C code used in the classical simulations. J.J.P. provided estimates of the quadrupole interaction strength. V.M., S.A., C.H., G.J.M., J.C.M., and A.M. conceived the experimental design and measurements proposed in this paper. V.M., S.A., and A.M. wrote the paper, with contributions from all authors. A.M. supervised the project.

\appendix

\section{Classical and quantum Hamiltonian parameters} \label{app: compare_H}

The classical simulation results shown in Fig.~\ref{fig: Figure 1} use dimensionless parameters which eases a direct comparison to corresponding dimensionless parameters in the quantum case. Here, their relation to the parameters in the original Hamiltonian [Eq.~\eqref{eq: H_top_classical}] and to the corresponding quantum parameters is given.

\subsubsection{Rescaling of classical Hamiltonian parameters}

Both the variables $\left(\mathbf{L}, t \right)$ and the parameters $\left(\alpha, \beta, \gamma, f \right)$ need a dimensionless equivalent. 
Dividing the classical Hamiltonian $\mathcal{H}_\mathrm{c}$ by $|\mathbf{L}|$ (including a factor $2\pi$ to convert from \si{\radian \per \second} to \si{\hertz}) we obtain
\begin{align}
\frac{\mathcal{H}_\mathrm{c}}{2 \pi |\mathbf{L}|} &= \frac{\alpha}{2 \pi}\frac{L_\mathrm{z}}{|\mathbf{L}|}+\frac{\beta}{2 \pi} \frac{L_\mathrm{x}^2}{|\mathbf{L}|}+\frac{\gamma}{2 \pi} \cos \left(2 \pi f t \right) \frac{L_\mathrm{y}}{|\mathbf{L}|} \nonumber \\
&=\frac{\alpha}{2 \pi} L_\mathrm{z}'+\frac{\beta}{2 \pi}  |\mathbf{L}|{L_\mathrm{x}'}^2+\frac{\gamma}{2 \pi} \cos \left(2 \pi f t \right) L_\mathrm{y}',
\end{align}
where the normalized angular momentum variable $\mathbf{L}'=\mathbf{L}/|\mathbf{L}|$ is introduced. Next we divide by $\alpha/2 \pi$ and time $t' = (\alpha / 2 \pi ) t$ is introduced:
\begin{align}
\mathcal{H}_\mathrm{c}' = \frac{\mathcal{H}_\mathrm{c}}{\alpha |\mathbf{L}|} &=L_\mathrm{z}'+\frac{\beta |\mathbf{L}|}{\alpha}{L_\mathrm{x}'}^2+\frac{\gamma}{\alpha}\cos \left(2 \pi \frac{2 \pi f }{\alpha} t'\right) L_\mathrm{y}' \nonumber \\
&=L_z'+\beta' {L_\mathrm{x}'}^2+\gamma'\cos \left( 2 \pi f' t' \right) L_\mathrm{y}'
\end{align}
This makes the parameters $\beta'$, $\gamma'$ and $f'$ dimensionless and relative to $\alpha$. $\alpha$ itself has units of \si{\radian \per \second} and time variable $t'$ has units of $2 \pi / \alpha$. 

\subsubsection{Correspondence between classical and quantum Hamiltonian parameters}

The above implies that the same convention of a normalized angular momentum has to be followed quantum mechanically. 
Hence the quantum Hamiltonian $\mathcal{H}_\mathrm{q}$ is divided by $h$ and transformed with $\mathbf{I'}=\mathbf{I}/I$, thus converting to frequency units and normalizing the spin operators:
\begin{align}
\frac{\mathcal{H}_\mathrm{q}}{h I}= \gamma_\mathrm{n} B_0 I_\mathrm{z}' + QI {I_\mathrm{x}'}^2 + \gamma_\mathrm{n} B_1 \cos \left( 2 \pi f t \right) I_\mathrm{y}',
\end{align}
This assumes units of \si{\hertz \per \tesla} for $\gamma_\mathrm{n}$, units of \si{\hertz} for $Q$, and dimensionless spin operators $\mathbf{I}$. 
Note that after introducing $\mathbf{I}'$ an additional factor $I$ appears in the second term due to its quadratic nature. 

Next, as in the classical case, dividing by $\gamma_\mathrm{n} B_0$ and introducing time variable $t'= \gamma_\mathrm{n} B_0 t$ results in the dimensionless Hamiltonian
\begin{align}
\mathcal{H}'_\mathrm{q}=\frac{\mathcal{H}_\mathrm{q}}{h I \gamma_\mathrm{n} B_0} &= I_\mathrm{z}'+\frac{QI}{\gamma_\mathrm{n}B_0}{I_\mathrm{x}'}^2 + \frac{B_1}{B_0} \cos \left(\frac{2 \pi f}{\gamma_\mathrm{n}B_0}t' \right) I_\mathrm{y}' \nonumber \\
&= I_\mathrm{z}'+Q'{I_\mathrm{x}'}^2+B_1'\cos \left(2 \pi f' t' \right) I_\mathrm{y}'
\end{align}

Table~\ref{tab: parameter_comparison} gives an overview of the different parameters used throughout the classical and quantum simulations. 

\begin{table} 
\caption{\label{tab: classical_quantum_comparison} Comparison of equivalent classical and quantum Hamiltonian parameters.}
\label{tab: parameter_comparison}
\begin{ruledtabular}
\begin{tabular}{cccc}
\multicolumn{2}{c}{\textbf{classical}} & \multicolumn{2}{c}{\textbf{quantum}} \\
original & dimensionless & original & dimensionless \\
$\mathbf{L}$ & $\mathbf{L}'=\frac{\mathbf{L}}{|\mathbf{L}|}$ & $\mathbf{I}$ & $\mathbf{I}'=\frac{\mathbf{I}}{I}$\\
$\frac{\alpha}{2 \pi}$ & 1 & $\gamma_\mathrm{n}B_0$ & 1 \\
$\frac{\beta}{2 \pi}$ & $\beta'=\frac{\beta |\mathbf{L}|}{\alpha}$ & $Q$ & $Q'=\frac{Q I}{\gamma_\mathrm{n}B_0}$ \\
$\frac{\gamma}{2 \pi}$ & $\gamma'=\frac{\gamma}{\alpha}$ & $\gamma_\mathrm{n} B_1$ & $B_1'=\frac{B_1}{B_0}$ \\
$f$ & $f'=\frac{2 \pi f}{\alpha}$ & $f$ & $f'=\frac{f}{\gamma_\mathrm{n}B_0}$ \\
$t$ & $t'=\frac{\alpha}{2 \pi} t$ & $t$ & $t'=\gamma_\mathrm{n}B_0 t$
\end{tabular}
\end{ruledtabular}
\end{table}

\section{Derivation of classical equations of motion} \label{app: clas_eq_of_motion}

The general expressions for the equations of motion of the driven top are given here. 
The starting point is Hamilton's equation of motion in Poisson bracket formulation:
\begin{equation} \label{eq: EOM_general_1}
\frac{\mathrm{d}\mathbf{L}}{\mathrm{d}t}=\{\mathbf{L},\mathcal{H}\}+\frac{\partial \mathbf{L}}{\partial t}
\end{equation}
with $\mathbf{L}=\left(L_\mathrm{x} \quad L_\mathrm{y} \quad L_\mathrm{z}\right)^\mathrm{T}$ the angular momentum vector, $\mathcal{H}$ the Hamiltonian, and $\{\mathbf{L},\mathcal{H}\}$ the Poisson bracket relation between $\mathbf{L}$ and $\mathcal{H}$. 
Angular momentum conservation implies $\partial \mathbf{L}/\partial t = 0$. 
Introducing $\dot{L}_\mathrm{i} \equiv \mathrm{d}L_\mathrm{i}/\mathrm{d}t$, the equations of motion are
\begin{eqnarray} \label{eq: EOM_general_2}
\dot{L}_\mathrm{x} =& \{L_\mathrm{x},\mathcal{H}\} \nonumber \\
\dot{L}_\mathrm{y} =& \{L_\mathrm{y},\mathcal{H}\} \\
\dot{L}_\mathrm{z} =& \{L_\mathrm{z},\mathcal{H}\} \nonumber
\end{eqnarray}

\subsection{Equations of motion for the classical driven top}
The equations of motion given above can be applied to the classical driven top Hamiltonian of the main text. 
As an example, we derive an expression for $\dot{L_\mathrm{y}}$:
\begin{eqnarray} \label{eq: }
\dot{L}_\mathrm{y} =& \{L_\mathrm{y},\alpha L_\mathrm{z}+\beta L_\mathrm{x}^2+ \gamma L_\mathrm{y} \cos \left(2 \pi f t \right) \}\nonumber \\
=& \alpha \underbrace{\{L_\mathrm{y},L_\mathrm{z}\}}_{= L_\mathrm{x}}+\beta \{L_\mathrm{y},L_\mathrm{x}^2\}+\gamma \cos \left(2 \pi f t \right) \underbrace{\{ L_\mathrm{y}, L_\mathrm{y}\}}_{=0} \nonumber \\
=& \alpha L_\mathrm{x}-\beta \underbrace{\{L_\mathrm{x},L_\mathrm{y}\}}_{=L_\mathrm{z}} L_\mathrm{x}-\beta L_\mathrm{x}\underbrace{\{L_\mathrm{x},L_\mathrm{y}\}}_{=L_\mathrm{z}}\\
=& \alpha L_\mathrm{x} - 2 \beta L_\mathrm{x} L_\mathrm{z} \nonumber
\end{eqnarray}
where the product rule for Poisson brackets is used in the third line and whenever Poisson brackets are computed, the relation $\{L_\mathrm{i},L_\mathrm{j}\}=\epsilon_\mathrm{ijk}L_\mathrm{k}$, with $\epsilon_\mathrm{ijk}$ the Levi-Civita symbol, is used. 
Similarly, equations for $L_\mathrm{x}$ and $L_\mathrm{z}$ can be derived, resulting in the system of equations
\begin{eqnarray}
\dot{L}_\mathrm{x} =& -\alpha L_\mathrm{y} + \gamma L_\mathrm{z} \cos \left(2 \pi f t \right) \nonumber\\
\dot{L}_\mathrm{y} =& \alpha L_\mathrm{x} - 2 \beta L_\mathrm{x} L_\mathrm{z} \\
\dot{L}_\mathrm{z} =& - 2 \beta L_\mathrm{x} L_\mathrm{y}-\gamma L_\mathrm{x} \cos \left(2 \pi f t \right) . \nonumber
\end{eqnarray}

\section{Quantum system in the rotating frame and rotating wave approximation} \label{app: RWA}

The technique of `dressing' a quantum spin state relies on applying a microwave tone with a frequency matching the dominant linear Zeeman interaction term in the system. 
Upon transforming the system to the rotating frame, and applying the rotating wave approximation (RWA), the original linear Zeeman interaction term disappears, and an effective linear interaction with a strength set by the microwave amplitude appears. 
We derive the effective ionized nuclear spin Hamiltonian of a donor in silicon using this approach; the starting point is the Hamiltonian proposed in Eq.~\eqref{eq: H_top_quantum_AM_mod},
\begin{align} \label{eq: H_quantum_RF_app}
& \mathcal{H}_\mathrm{quantum,RF} = \gamma_\mathrm{n}B_\mathrm{0} I_\mathrm{z}+Q I_\mathrm{x}^2 + \\ 
& \left(\gamma_\mathrm{n}B_\mathrm{1,I} \cos{\left(2\pi f_\mathrm{RF}t\right)}
+\gamma_\mathrm{n}B_\mathrm{1,Q}\cos{\left(2\pi f t \right)} \sin{\left(2\pi f_\mathrm{RF} t \right)}\right)I_\mathrm{y} \nonumber
\end{align}

\subsection{Transforming spin operators to the rotating frame}

The transformation of $\mathcal{H}_\mathrm{quantum,RF}$ to a frame rotating with angular velocity $\omega_\mathrm{RF}=2\pi f_\mathrm{RF}$ is given by
\begin{equation}
\mathcal{H}_\mathrm{quantum,RF}'= R\left(-\omega_\mathrm{RF}t\right)\mathcal{H}_\mathrm{quantum,RF} R\left(\omega_\mathrm{RF} t \right)-f_\mathrm{RF} I_\mathrm{z}
\end{equation}
where $R$ is the rotation operator corresponding to a basis rotation over an angle $\phi$ around the $z$ axis:
\begin{equation}
R\left(\phi\right) = e^{-i\phi I_\mathrm{z}} = e^{-i\omega_\mathrm{RF}  t I_\mathrm{z}}
\end{equation}
The remaining task is to transform the (squared) spin operators of the original Hamiltonian to the rotating frame. 
Using the series expansion of the matrix exponent, commutator rules of spin operators and recognizing sine or cosine series in the expansion, one can derive the following identities for rotated spin operators:
\begin{align}
\nonumber
R\left(-\phi\right)I_\mathrm{z}R\left(\phi\right) &= I_\mathrm{z} \\
R\left(-\phi\right)I_\mathrm{y}R\left(\phi\right) &= \sin{\phi} I_\mathrm{x} + \cos{\phi} I_\mathrm{y}\\ \nonumber
R\left(-\phi\right)I_\mathrm{x}R\left(\phi\right) &= \cos{\phi} I_\mathrm{x} - \sin{\phi} I_\mathrm{y}\\ \nonumber
\end{align}
With the same strategy, albeit less trivially, one can derive for the squared spin operators the following identities:
\begin{align}
\nonumber
R\left(-\phi\right)I_\mathrm{z}^2R\left(\phi\right) &= I_\mathrm{z}^2 \\
R\left(-\phi\right)I_\mathrm{y}^2 R\left(\phi\right) &= -\frac{1}{2}I_\mathrm{z}^2+\frac{1}{2}\sin{2\phi}\{I_\mathrm{x},I_\mathrm{y}\} \\ \nonumber 
&-\frac{1}{2}\cos{2\phi}\left(I_\mathrm{x}^2-I_\mathrm{y}^2\right)+\frac{1}{2}I\left(I+1\right)\\ \nonumber
R\left(-\phi\right)I_\mathrm{x}^2 R\left(\phi\right) &=  -\frac{1}{2}I_\mathrm{z}^2 -\frac{1}{2}\sin{2\phi}\{I_\mathrm{x},I_\mathrm{y}\} \\ \nonumber 
&+\frac{1}{2}\cos{2\phi}\left(I_\mathrm{x}^2-I_\mathrm{y}^2\right)+\frac{1}{2}I\left(I+1\right)\\ \nonumber
\end{align}
where $\{I_\mathrm{x},I_\mathrm{y}\}$ denotes the anticommutator of $I_\mathrm{x}$ and $I_\mathrm{y}$.

\subsection{Hamiltonian under the rotating wave approximation}

Using the results of the previous section, it is straightforward to arrive at the rotating frame version of Eq.~\ref{eq: H_top_quantum_AM_mod}, which is given by:
\begin{align}
&\mathcal{H}_\mathrm{quantum,RF}'= -\frac{1}{2}QI_\mathrm{z}^2 -\frac{1}{2}Q\sin{2\omega_\mathrm{RF} t}\{I_\mathrm{x},I_\mathrm{y}\}\\ \nonumber 
&+\frac{1}{2}Q\cos{2\omega_\mathrm{RF} t}(I_\mathrm{x}^2-I_\mathrm{y}^2)+\frac{1}{2}QI\left(I+1\right)\\ \nonumber
&+\frac{1}{2}\left(\gamma_\mathrm{n} B_\mathrm{1,I}\sin{2\omega_\mathrm{RF} t} + \gamma_\mathrm{n} B_\mathrm{1,Q}\cos{\omega t}(1+\cos{2 \omega_\mathrm{RF} t}) \right)I_\mathrm{x} \\ \nonumber 
&-\frac{1}{2}\left(\gamma_\mathrm{n} B_\mathrm{1,I}\left(1-\cos{2\omega_\mathrm{RF} t}\right) + \gamma_\mathrm{n} B_\mathrm{1,Q}\cos{\omega t}\sin{2 \omega_\mathrm{RF} t}\right)I_\mathrm{y}
\end{align}
with $\omega = 2 \pi f$ the drive frequency to create the quantum driven top, and $\omega_\mathrm{RF} = 2 \pi f_\mathrm{RF}$, with $f_\mathrm{RF}=\gamma_\mathrm{n}B_0$ the drive frequency of the rotating frame (exactly canceling the Zeeman interaction term $\gamma_\mathrm{n}B_0 I_\mathrm{z}$). 
Applying the RWA now reduces to neglecting all terms involving oscillatory factors at frequency $2\omega_\mathrm{RF}$. 
Further ignoring the irrelevant static energy offset $\dfrac{1}{2}QI\left(I+1\right)$, the Hamiltonian reduces to Eq.~\eqref{eq: H_quantum_top_RWA}:
\begin{align}
\mathcal{H}_\mathrm{quantum, RWA} = & - \frac{1}{2} \gamma_\mathrm{n} B_\mathrm{1,I}  I_\mathrm{y}\\ \nonumber
& -\frac{1}{2}QI_\mathrm{z}^2+\frac{1}{2}\gamma_\mathrm{n} B_\mathrm{1,Q}\cos{\left(2\pi f t\right)}I_\mathrm{x}
\end{align}
Upon applying two trivial rotations, first by an angle $\pi/2$ around the $y$ axis, followed by an angle $\pi/2$ around the $x$ axis, the original quantum driven top Hamiltonian of Eq.~\eqref{eq: H_top_quantum} is recovered, demonstrating equivalence between the laboratory and rotating frame approach to creating a quantum driven top.

\subsection{Relative angle between quadrupole interaction and periodic drive}
The Hamiltonian of Eq.~\eqref{eq: H_quantum_RF_app} still contains strong constraints on the relative directions of the different terms, as the linear Zeeman interaction, quadratic quadrupole interaction, and periodic drive are all orthogonal to each other. 
In a realistic experiment, the angle $\theta$ between the principal axis of the quadrupole interaction and the direction of the periodic driving field will be an intrinsic and uncontrollable device property. 
Under the RWA, however, the constraint of orthogonality between quadrupole interaction and periodic drive may be relaxed. 
Defining the $x,y$ plane as the one containing the principal axis of the quadrupole coupling and the periodic driving field, the periodic drive term in Eq.~\eqref{eq: H_quantum_RF_app} may be rewritten as
\begin{align}
&[\gamma_\mathrm{n}B_\mathrm{1,I}\sin{\left(2\pi f_\mathrm{RF}t - \varphi \right)}+ \\ \nonumber
& -\gamma_\mathrm{n}B_\mathrm{1,Q}\cos{\left(2\pi f t \right)} \cos{\left(2\pi f_\mathrm{RF} t - \varphi \right)}]\left(\cos\theta I_\mathrm{x}+\sin \theta I_\mathrm{y}\right)
\end{align}

where a phase shift by an angle $\varphi$ is included in the drive. 
Upon choosing $\varphi$ to be equal to the angle $\theta$, in the rotating frame, the only oscillatory factors containing the phase $\theta$ are all terms in $2 \omega_\mathrm{RF}$, which are neglected under the RWA. 
One then recovers as the remaining terms the desired combination $-\tfrac{1}{2}\gamma_\mathrm{n}B_\mathrm{1,I}I_\mathrm{y}+\tfrac{1}{2}\gamma_\mathrm{n}B_\mathrm{1,Q}\cos\left( 2\pi f\right) I_\mathrm{x}$. 
This implies that one may correct for the angle $\theta$ between quadrupole interaction and drive axis straightforwardly by including this angle as a phase shift in the periodic drive. 

\subsection{Classical chaos for rotating frame parameters}
\begin{figure}
\includegraphics[width=\columnwidth]{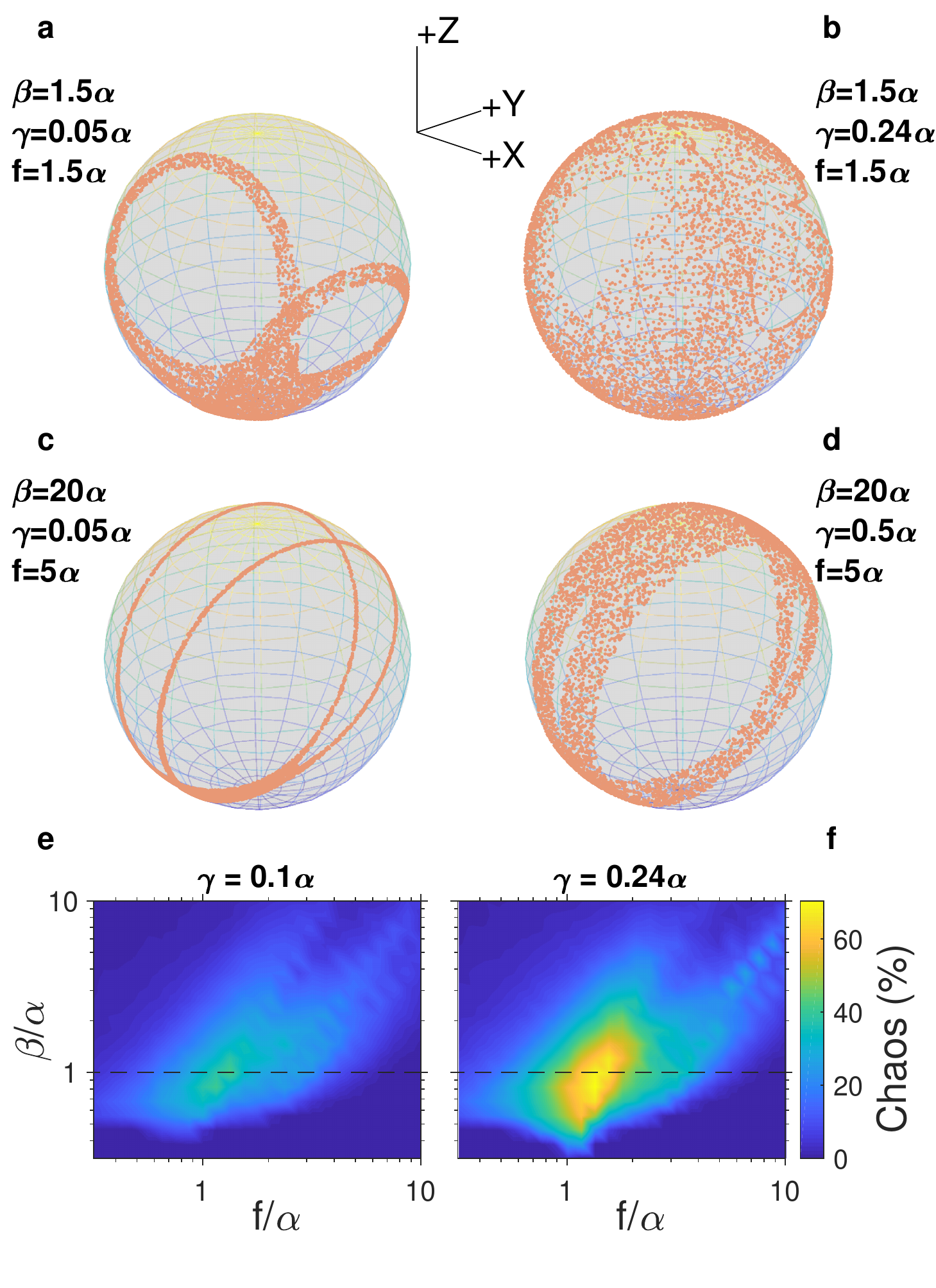}
\caption{
Classical chaos for large quadratic interaction strength and strong periodic drive.
These regimes are particularly relevant to the rotating-frame implementation of the quantum driven top. 
(a),(b) Quadratic interaction strength $\beta = 1.5 \alpha$, a regime very similar to Fig.~\ref{fig: Figure 1}(c). 
In (a) the periodic drive strength $\gamma=0.05\alpha$ is very similar to the drive strength in Fig.~\ref{fig: Figure 1}(e). 
In (b), strong periodic drive $\gamma = 0.24 \alpha$ is considered, showing very large chaotic phase space fractions. 
For even larger strengths of $\gamma$, the whole phase space is chaotic. 
(c),(d) Dominant quadratic interaction strength $\beta = 20 \alpha$. 
In (c), $\gamma = 0.05 \alpha$, and a small chaotic area is present in the phase space. 
In (d), $\gamma = 0.5 \alpha$, and a sizable fraction of the phase space is chaotic. 
(e),(f) show the chaotic fraction of the total phase space as a function of $\beta$ and $f$, for two different periodic drive strengths $\gamma = 0.1 \alpha$ (e) and $\gamma = 0.24 \alpha$ (f). 
This clearly demonstrates how large chaotic fractions of the phase space are still attainable even with $\beta$ large compared to $\alpha$.
}
\label{fig: app_Classical_chaos}
\end{figure}

As mentioned in the main text, by working in the rotating frame new parameters regimes may be explored that are difficult to reach in the laboratory frame. 
In particular, the regime of quadrupolar interaction strength much larger than effective linear interaction strength may now be explored. 
Similarly, the regime of large periodic drive may be explored, as the technique introduced here allows for any periodic drive strength up to the effective linear interaction strength. 
We extended our analysis of the classical dynamics presented in the main text to explore these regimes as well. 
Not surprisingly, we find that the regime of large periodic drive strength allows for very large chaotic fractions of the classical phase space. Even in regimes where the linear interaction strength is an order of magnitude smaller compared to the quadratic interaction strength, we still find significant chaotic fractions.
These findings are summarized in Fig.~\ref{fig: app_Classical_chaos}. 

\section{Quantum dynamics and Floquet formalism} \label{app: floquet}

The evolution of time-dependent periodic Hamiltonians, such as the quantum driven top, can be transformed to time-independent evolutions using the Floquet formalism.
The Floquet operator $\mathcal{F}$ is equal to the time evolution operator $\mathcal{U}$ over one full period $\tau$:

\begin{equation}
\mathcal{F} \equiv \mathcal{U}(\tau, 0) = \mathbbm{1}- i/\hbar\int_0^\tau \mathcal{H}\left(t'\right)\mathcal{U}\left(t',0\right)dt',
\end{equation}

As a result, the Floquet operator $\mathcal{F}$ has the property $\ket{\psi(\tau)} = \mathcal{F} \ket{\psi(0)}$, irrespective of the state $\ket{\psi(0)}$.
A consequence is that $\ket{\psi(N \tau)} = \mathcal{F}^N \ket{\psi(0)}$, and so once $\mathcal{F}$ is known, any state can be straightforwardly evolved over a discrete number of periods by repeated application of $\mathcal{F}$.

The Floquet operator can be decomposed into eigenstates $\ket{\Phi_i}$ and corresponding eigenvalues $\lambda_i$, which all satisfy $\left|\lambda_i\right|=1$ ($\mathcal{F}$ is unitary).
All eigenvalues are therefore of the form $\lambda_i=\exp{\left(-i \epsilon_i \tau /\hbar\right)}$, where the angular frequency $\epsilon_i$ is known as the quasienergy of the corresponding eigenstate.
This enables decomposition of any initial state $\ket{\psi(0)}$ into the Floquet eigenstates, and straightforward calculation of its state after evolution of $N$ periods:

\begin{align}
\ket{\psi(N\tau)} = & \mathcal{F}^N \ket{\psi(0)} \nonumber \\
	= & \sum_i \braket{\Phi_i|\psi(0)}\mathcal{F}^N\ket{\Phi_i}  \\
    = & \sum_i \braket{\Phi_i|\psi(0)} \exp{\left( -i \epsilon_i N \tau/\hbar\right)}\ket{\Phi_i} \nonumber
\end{align}

Each Floquet eigenstate accumulates a phase determined by its respective quasienergy, and so superpositions of eigenstates will lead to interference effects.
This offers an explanation for the evolution of spin coherent states (Fig.~\ref{fig: figure 4}); whereas the spin coherent state with many Floquet components [orange  (light gray)] has many interfering frequencies, the state that exhibits dynamical tunneling [blue (dark gray)] primarily consists of only two Floquet components. 
In the latter case, there are two main interfering frequencies, and their difference determines the dynamical tunneling frequency.
The Floquet formalism also connects to the varying dephasing rates for different initial states (Fig.~\ref{fig: figure 3}), as the individual Floquet quasienergies have a different degree of sensitivity to perturbations.
More generally, the Floquet formalism clearly emphasizes the discrete nature of quantum mechanics, leading to quasiperiodicity that causes partial revivals of initially localized states, as opposed to the exponential divergence of trajectories in the case of classical chaos. 

\section{Spin $I>1/2$: some properties and definitions} \label{app: spin_properties}

\subsection{Spin coherent states}

Spin coherent states, also known as Bloch states, form a subset of the possible states of a spin-$I$ system.
The formulation and some relevant properties of the spin coherent states are reviewed here.

For a spin with spin quantum number $I$ and magnetic quantum number $m \in \left[-I, -I+1, ..., I-1, I\right]$, states can be described by the basis states $\ket{I,m}$, which are eigenstates of the $I_\mathrm{z}$ spin operator with corresponding eigenvalue $m$.

We introduce spherical angles $\phi$ (azimuthal) and $\theta$ (polar), and an operator $R_\mathrm{\theta,\phi}$ corresponding to a rotation over an angle $\theta$ about an axis $\left(\sin\phi, -\cos\phi,0\right)$, given by
\begin{equation}
R_\mathrm{\theta,\phi}=e^{-i \theta \left(I_\mathrm{x} \sin \phi -I_\mathrm{y} \cos \phi \right)}
\end{equation} 
The spin coherent states $\ket{\theta,\phi}$ are now defined as the state $\ket{I,I}$ rotated by $R_\mathrm{\theta,\phi}$~\cite{arecchi1972coherentstates}:
\begin{align}
\ket{\theta,\phi} = & R_\mathrm{\theta,\phi}\ket{I,I}\nonumber \\
= & \sum_{m = -I} ^{I} \binom {2I}{I+m} ^{1/2} e^{i\phi\left(I-m\right)} \, \times \, \ldots \\
& \ldots \, \times \, \left(\cos\tfrac{1}{2}\theta\right)^{I+m}\left(\sin\tfrac{1}{2}\theta\right)^{I-m}\ket{I,m} \nonumber 
\end{align}

All spin coherent states share the property
\begin{equation}
|\mathbf{I}|=\sqrt{\braket{I_\mathrm{x}}^2+\braket{I_\mathrm{y}}^2+\braket{I_\mathrm{z}}^2} = I
\end{equation}
Furthermore, the spin coherent states form an overcomplete normalized basis, having nonzero overlap with each other. Only opposite spin coherent states with a $\pi$ difference in polar angle $\theta$ are orthogonal and have zero overlap, analogous to the orthogonality of $\ket{I,-I}$ and $\ket{I,I}$. 

Rotated spin operators can be obtained by applying $R_\mathrm{\theta,\phi}$ to the original spin operators:
\begin{align}
I_\mathrm{x'}=R_\mathrm{\theta,\phi}I_\mathrm{x}R_\mathrm{\theta,\phi}^{-1}, \nonumber \\ I_\mathrm{y'}=R_\mathrm{\theta,\phi}I_\mathrm{y}R_\mathrm{\theta,\phi}^{-1},\\ I_\mathrm{z'}=R_\mathrm{\theta,\phi}I_\mathrm{z}R_\mathrm{\theta,\phi}^{-1}, \nonumber
\end{align}
in which case the spin coherent state $\ket{\theta,\phi}$ is an eigenstate of $I_\mathrm{z'}$ with eigenvalue $I$. 
Spin coherent states are the only states for which the uncertainty relation ($\sigma_{I_\mathrm{x'}}^2\sigma_{I_\mathrm{y'}}^2\geq\tfrac{\hbar^2}{4}\braket{I_\mathrm{z'}}^2$) becomes an equality, and are therefore also known as minimum-uncertainty states.

For $I = 1/2$, all pure spin states are spin coherent states, leaving the concept unnecessary. 
However, this concept is very useful for $I > 1/2$, since now the spin coherent states form a distinct subset of pure states for which the spin is maximally aligned in a certain direction $(\theta,\phi)$, with a minimum-uncertainty spread around it.
As such, these states are the closest analog to classical angular momentum, and are therefore used in the quantum driven-top experiments as corresponding initial quantum states.

\subsection{Husimi Q distribution}

\begin{figure}
\includegraphics[width=\columnwidth]{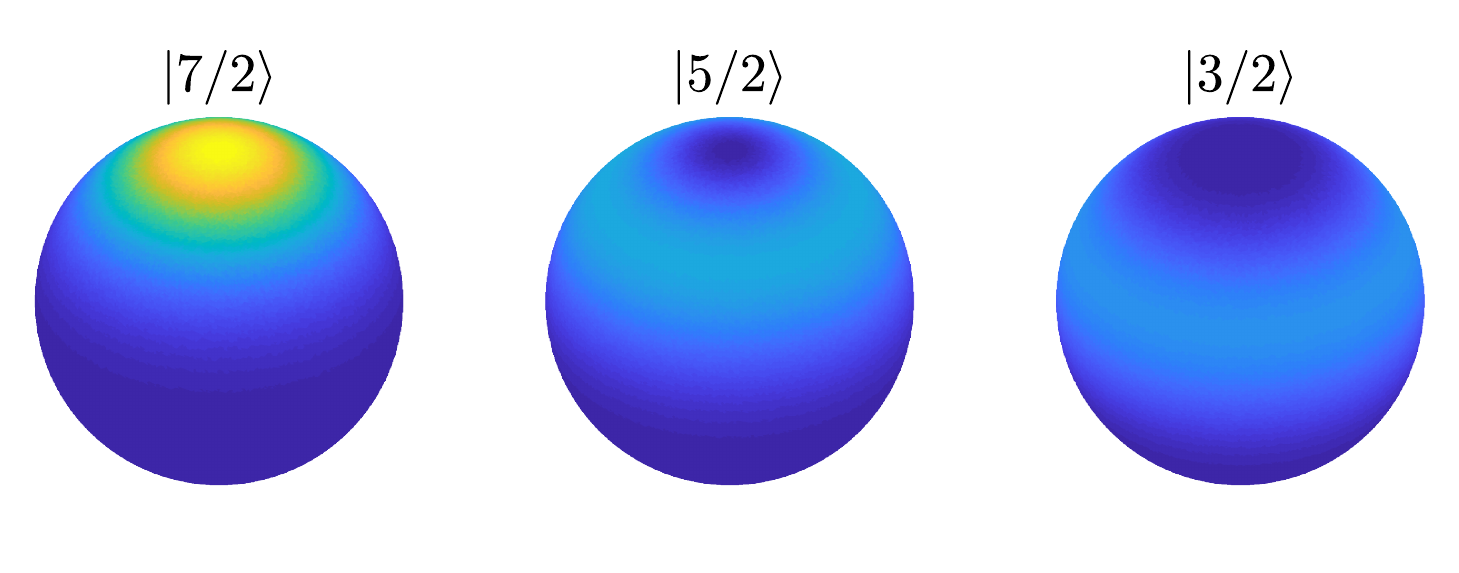}
\caption{Husimi representation of $I_z$ eigenstates.
Colors on the sphere surface correspond to the Husimi Q function [Eq.~\eqref{eq: Husimi Q function}] evaluated at those spherical coordinates.
Color-scale limits are constant over all panels, varying between $0$ [dark blue (dark gray)] and $1/\pi$ [bright yellow (light gray)]. 
Three eigenstates of $I_\mathrm{z}$ are visualized for $I=7/2$.
Whereas the state $\ket{7/2}$ is a spin coherent state oriented along $+z$ with minimum uncertainty, as can be seen by its small spread over the spherical surface, the other two eigenstates are uniform bands with a larger spread since these are not minimum-uncertainty states.
}
\label{fig: figureAppHusimiEigenstates}
\end{figure}

\begin{figure}
\includegraphics[width=\columnwidth]{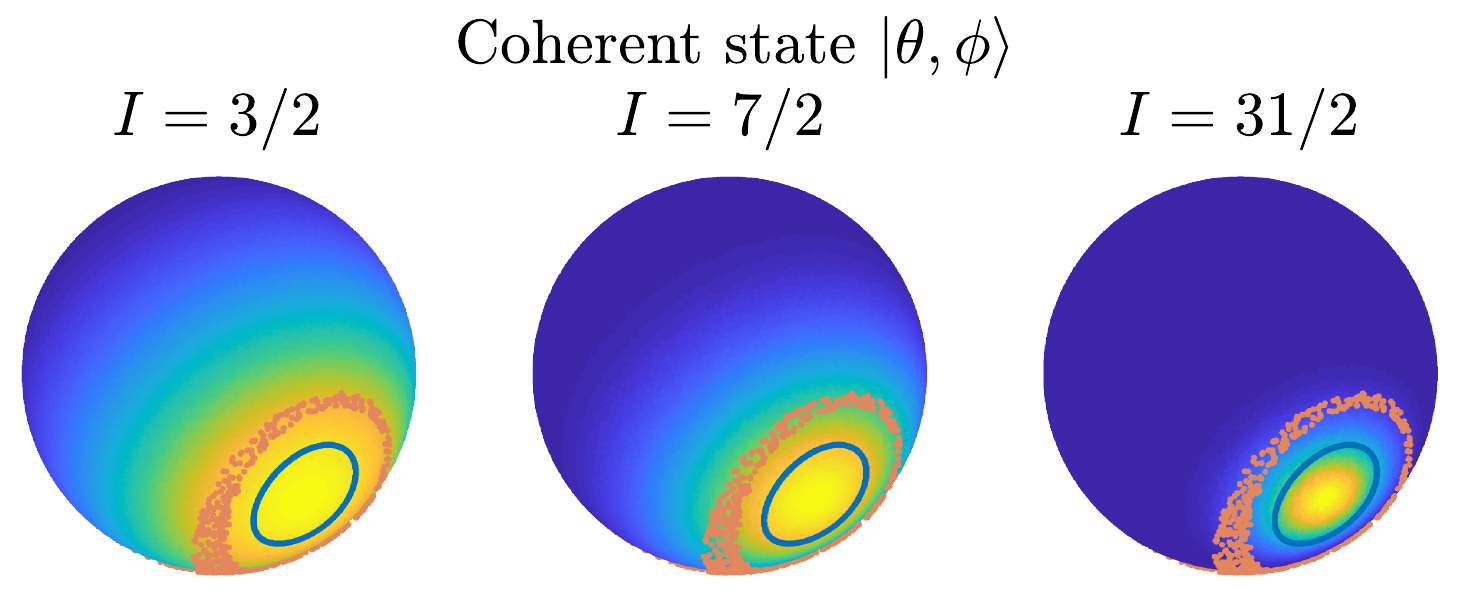}
\caption{
Localization of a spin-coherent state within regions of classical phase space.
Colors of the sphere surface correspond to the Husimi Q function [Eq.~\eqref{eq: Husimi Q function}] of the coherent state $\ket{\theta , \phi}=\ket{0.675 \pi, 0.475 \pi}$ evaluated at those spherical coordinates.
Color-scale limits are constant over all panels, varying between $0$ [dark blue (dark gray)] and $1/\pi$ [bright yellow (light gray)]. 
Stroboscopic maps of two trajectories of the classical driven top are superimposed on the same spherical surface, with the same parameters as chosen in Fig.~\ref{fig: Figure 2}(a). 
The chaotic trajectory [orange (light gray dots)] encloses an island of stability in which the regular trajectory [blue (dark gray dots)] resides. 
As $I$ is increased from 3/2 (left panel, e.g., an \As nuclear spin) to 7/2 (middle panel, e.g., an \Sb nuclear spin) and beyond (right panel, 31/2, shown for pedagogical reasons only, not corresponding to an actual nuclear spin), the relative uncertainty of the coherent state decreases, effectively localizing the state within the different regions of the corresponding classical phase space.
This underlines the importance of choosing a donor with a large nuclear spin for a meaningful comparison between the dynamics of quantum states and the classically regular or chaotic counterpart.
}
\label{fig: figurevaryI}
\end{figure}

The Husimi Q distribution~\cite{husimi1940} is a quasiprobability distribution that is used to represent quantum states.
It is particularly useful here to provide a visualization of high-dimensional quantum states (Fig.~\ref{fig: figureAppHusimiEigenstates}).
For a given density matrix $\rho$, the Husimi Q function is defined as
\begin{equation}
Q(\theta, \phi) = \frac{1}{\pi}\braket{\theta, \phi | \rho | \theta, \phi},
\label{eq: Husimi Q function}
\end{equation}
where $\ket{\theta, \phi}$ is a spin coherent state.
In the case of a pure state $\left(\rho=\ket{\psi}\bra{\psi}\right)$, the Husimi Q distribution simplifies to $Q(\theta, \phi) = \tfrac{1}{\pi}\left|\braket{\psi|\theta, \phi}\right|^2$, the overlap-squared between $\ket{\psi}$ and $\ket{\theta, \phi}$.

The Husimi Q distribution satisfies certain properties required for a joint probability distribution, because the distribution is normalized and non-negative with values ranging between $0 \leq Q(\theta, \phi) \leq 1/\pi$.
However, since different spin coherent states are nonorthogonal, different coordinates $(\theta,\phi)$ do not represent distinct physical contingencies, and different values of $Q$ are not the probability of mutually exclusive states, a requirement for a joint probability distribution.
This reflects the fact that quantum mechanics lacks a clear phase-space description, as opposed to classical mechanics.
Although not providing a true phase-space description, quasiprobability distributions such as the Husimi Q distribution provide the closest quantum-mechanical proxy to it as an alternative and complete representation of a quantum state (invertible to the original density matrix representation). 

The Husimi Q distribution is preferred here over the Wigner distribution as the phase space representation of quantum states because classical effects are emphasized.
For instance, application of the Husimi Q function to a spin coherent state closely matches a classical point in phase space with the addition of an uncertainty spread (Fig.~\ref{fig: figurevaryI}). 
A downside is that typical quantum phenomena such as interference are not clearly visible.
To visualize these effects, other distributions such as the Wigner distribution are more attractive candidates.

\subsection{Uncertainty and size of $I$}

Spin coherent states are the set of minimum uncertainty states ($\sigma_{I_\mathrm{x'}}^2\sigma_{I_\mathrm{y'}}^2 = \tfrac{\hbar^2}{4}\braket{I_\mathrm{z'}}^2 = \tfrac{\hbar^2}{4}I^2$). 
Its uncertainty defines a typical area over which a coherent state spreads in phase space, and should be compared to the total surface of the phase space ($4 \pi I^2$). 
This leads to a concept of relative uncertainty $\sigma_{I_\mathrm{x'}}\sigma_{I_\mathrm{y'}}/4 \pi I^2 \propto 1/I$, which is a measure of how well a quantum state is localized in phase space. 
This underlines the importance of having large enough $I$, as illustrated by comparing the Husimi Q function of spin coherent states for different $I$ (Fig.~\ref{fig: figurevaryI}).

\begin{figure}
\includegraphics[width=\columnwidth]{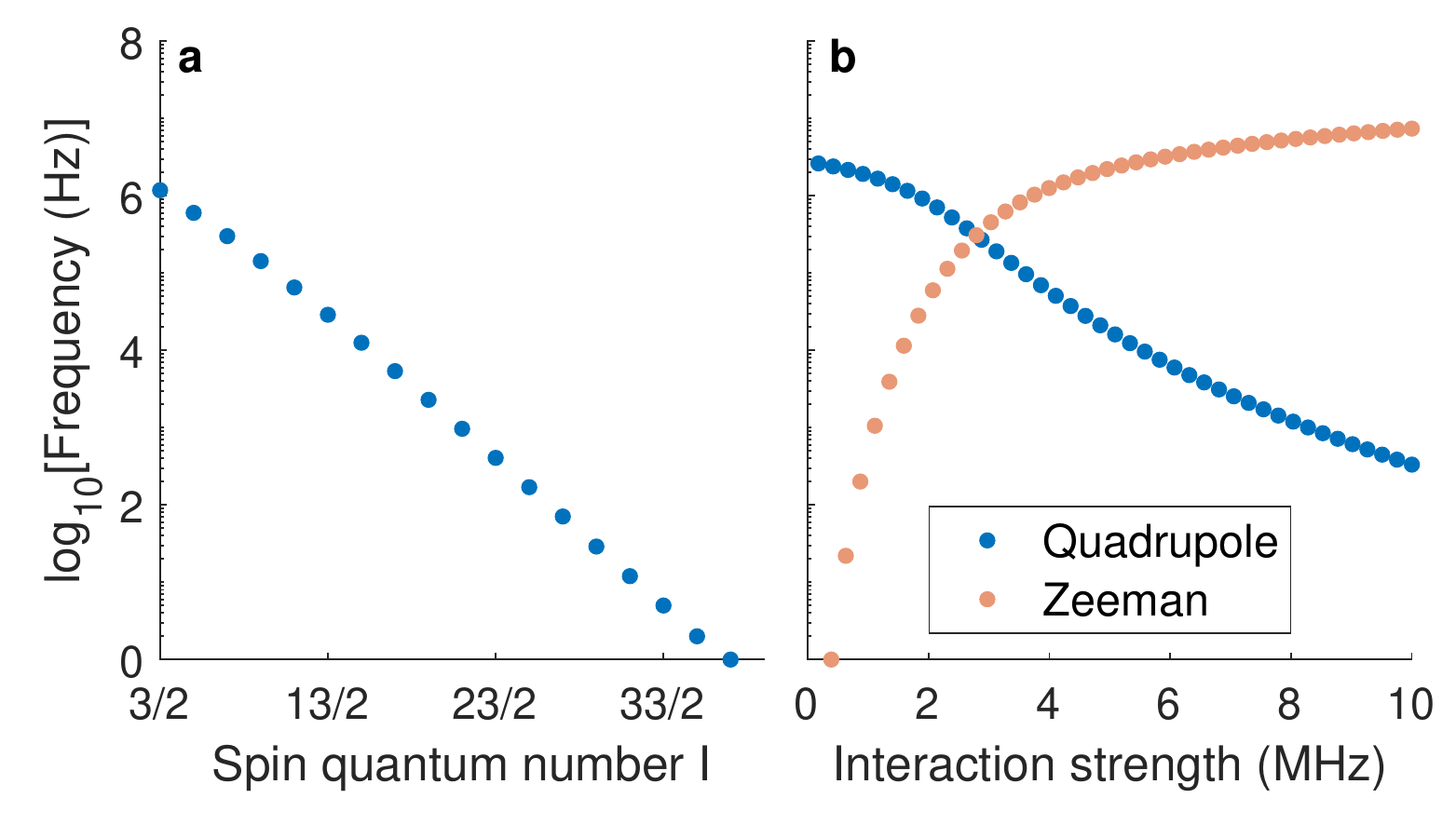}
\caption{
Dynamical tunneling frequency for varying system parameters.
(a) Dynamical tunneling frequency versus spin quantum number $I$. 
The static magnetic field $B_0=0.5$~T is kept constant and the drive is turned off ($B_1=0$~T), while the quadrupole strength is scaled such that $Q I = 2.8$~MHz to ensure equal linear and quadratic contribution (see Appendix~\ref{app: compare_H}). 
The dynamical tunneling frequency decreases exponentially with increasing $I$.
(b) Dynamical tunneling frequency for varying quadrupole interaction strength [blue (dark gray)] and Zeeman interaction strength [orange (light gray)].
The spin quantum number is fixed at $I=7/2$, $B_1 = 0$~T, and $-\gamma_n B_0 = 2.8$~MHz for varying quadrupole interaction $Q I$ [blue (dark gray)], and $Q I = 2.8$~MHz for varying Zeeman interaction $-\gamma_n B_0$ [orange (light gray)].
}
\label{fig: figureAppDynamicalTunnelingDependency}
\end{figure}

\section{Dependence of dynamical tunneling rate on system parameters} \label{app: dynamical tunneling}

We investigate how the dynamical tunneling rate is influenced by different system parameters.
Dynamical tunneling arises naturally in the quantum system, since the uncertainty spread of a quantum state prevents it from being truly localized within a classical region of phase space. 
Therefore, a state prepared within one of the classically stable regions of the driven top will have a finite overlap of its wave function with the other classically stable region, and may tunnel back and forth between these two regions. 
Qualitatively, the dynamical tunneling rate can be understood as determined by the amount of such a wave function overlap.

To get a better understanding of the tunneling rate we have numerically studied its dependence on different system parameters.
First, as the spin quantum number $I$ grows, the relative uncertainty spread of a spin coherent state on  the sphere of radius $I$ shrinks, and the dynamical tunneling rate decreases accordingly [Fig.~\ref{fig: figureAppDynamicalTunnelingDependency}(a)].
Second, the parameters $B_0$ and $Q$ define the relative distance of the corresponding classically stable regions.
Upon increasing $B_0$, the two stable regions come together, and the tunneling rate increases [Fig.~\ref{fig: figureAppDynamicalTunnelingDependency}(b)], whereas increasing $Q$ has the opposite effect. 

\begin{figure}
\includegraphics[width=\columnwidth]{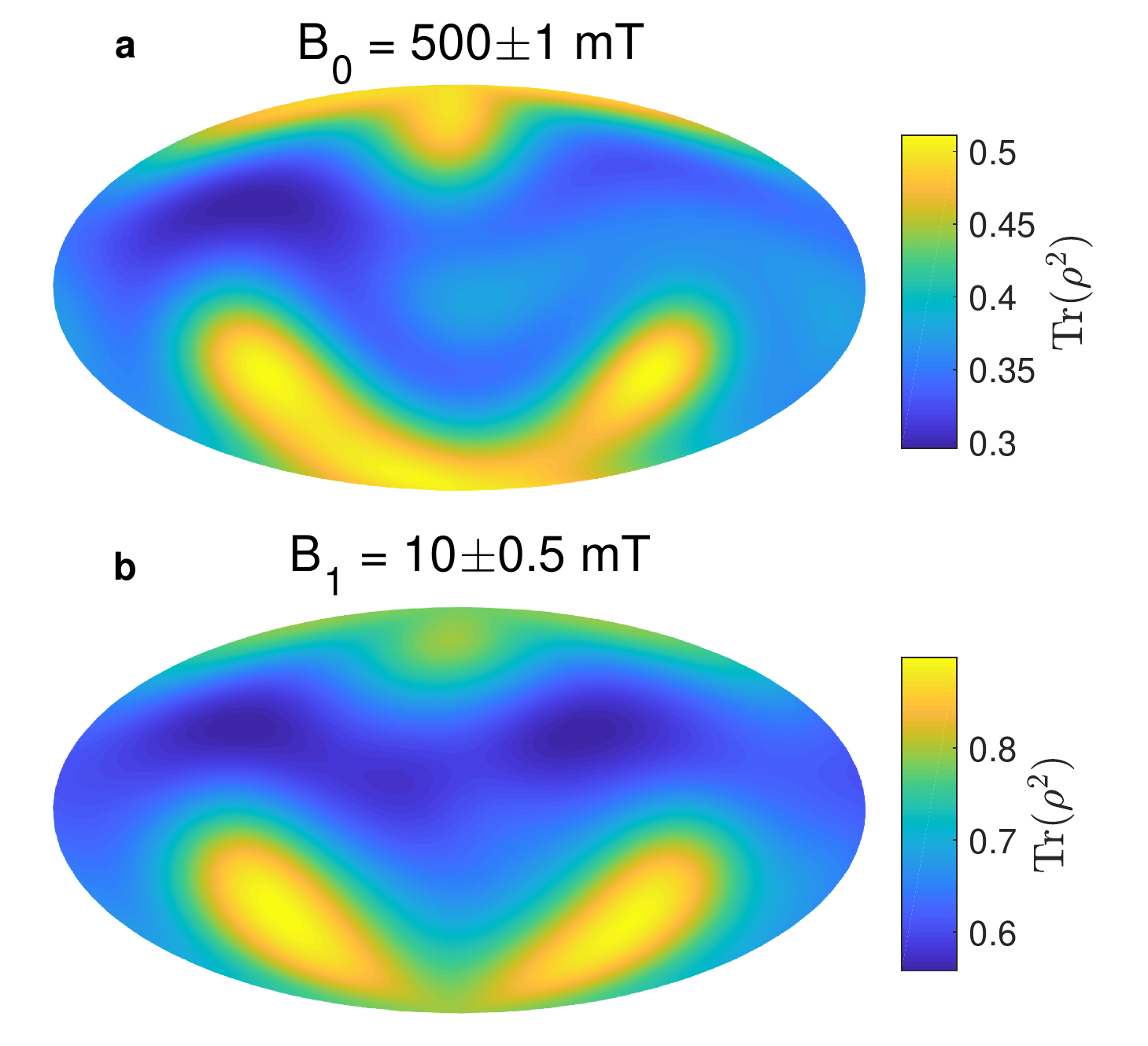}
\caption{
Quantum state purity of coherent states for fluctuating parameters $B_0$ and $B_1$.
The static magnetic field $B_0$ (a) and oscillating magnetic field $B_1$ (b) are varied once per period, keeping all other parameters fixed ($Q=800$~kHz, $B_0=0.5$~T, $B_1=0.01$~T, $A=0$~MHz, $f=3.5$~MHz).
Simulation details are identical to those in Fig.~\ref{fig: figure 3}(b), where the purity is shown for varying $Q$ (see Appendix{app: methods} for details).
The number of drive periods has been varied to account for varying sensitivity to perturbations ($N=2000$~periods for $B_0$, $N=10000$~periods for $B_1$).
In all cases, the resulting purity is qualitatively similar: the classical regular islands have a relatively high purity, and the classically chaotic areas surrounding these have a lower purity after the same evolution time.
}
\label{fig: FigureAppPurityOtherParameters}
\end{figure}

\section{Quantum state purity vs fluctuations in $B_0$ and $B_1$} \label{app: purity}

In addition to Fig.~\ref{fig: figure 3}, where fluctuations in the parameter $Q$ are considered, fluctuations in the parameters $B_0$ and $B_1$ are studied here (Fig.~\ref{fig: FigureAppPurityOtherParameters}). 
The approach used to obtain the results shown in Fig.~\ref{fig: figure 3} is repeated for this study and details of these simulations may be found in Appendix~\ref{app: methods}.
Similar behavior is found compared to considering fluctuations in $Q$.
We note that in all cases the correspondence between classical phase space being regular or chaotic and quantum state purity decaying slow or fast is not exact. Rather the general structure of the underlying classical phase space is recovered; i.e., classically stable points correspond to quantum states being less sensitive to fluctuations. 

\section{Nuclear quadrupole interaction} \label{app: quadrupole}

As pointed out in the main text, nuclei with a spin $I > 1/2$ have a quadrupole moment $Q_\mathrm{n}$ due to a nonspherical charge distribution~\cite{KauRMP79}. 
This charge distribution has an axis of symmetry that aligns with the nuclear angular momentum and interacts with an electric field gradient (EFG). 
We estimate the strength of quadrupole interaction due to such an EFG for the different donors in silicon. 

For group-V donors in silicon, the EFG is produced by external charges, such as the donor-bound electron or the crystal lattice. 
In a sample of unperturbed bulk silicon, the tetrahedral donor symmetry results in canceling EFG components and consequently a vanishing quadrupole interaction in the ground state. 
In a realistic micro- or nanoelectronic device, strain and/or electric fields can break the bulk symmetry of the donor wave functions~\cite{SliPMR90}. 
Strain is typically present in devices as a result of the mismatch of the coefficient of thermal expansion of different materials, for example between the metal electrodes and semiconductor substrate~\cite{ThoAIPA15, PlaarXiv16}. 
The subsequent rearrangement of the multivalley electron state generates electric field gradients which can couple to the electric quadrupole moment of the donor~\cite{PlaarXiv16, FraPRB16}. 
Strain also acts on the silicon crystal to produce an EFG through the gradient elastic tensor $\mathbf{S}$~\cite{SunPRB79}, which results in a quadrupole interaction even in the absence of the donor-bound electron \cite{FraPRL15}.

\subsection{Estimates of Nuclear Quadrupole Interaction}

We define the quadrupole interaction strength as the factor that precedes the quadratic $I_\mathrm{z'}$ term in Eq.~\eqref{eq: H_Q} of the main text,
\begin{equation}
\label{eq: Q_strength}
Q = \frac{3(1-\gamma_\mathrm{s}) e Q_\mathrm{n} V_{\rm z'z'}}{4I\left(2I-1\right)h}. 
\end{equation}
Here $\gamma_\mathrm{s}$ is the Sternheimer antishielding factor, $e$ is the elementary charge, $Q_\mathrm{n}$ is the nuclear quadrupole moment, $V_\mathrm{i j}=\tfrac{\partial^2 V}{\partial {i} \partial {j}}$, $\left(i,j \in x',y',z'\right)$, are the partial second derivatives of electric potential $V$, and $h$ is Planck's constant.
This differs from the more conventional definition of the quadrupole interaction (by including multiplication factor $3/4I\left(2I-1\right)$), as this allows a direct comparison to the equivalent classical parameter $\beta$. 
Accurate calculation of $Q$ for donors is complicated by the multiplicative term $\gamma_\mathrm{s}$. 
This factor relates to the Sternheimer antishielding effect, a phenomenon that describes the rearrangement of the inner electron shells in response to an external EFG, effectively enhancing the EFG experienced by the nucleus \cite{KauRMP79}. 
The Sternheimer antishielding factor $\gamma_\mathrm{s}$ can be considerable; theoretical calculations~\cite{FeiPR69} for isolated As and Bi ions show an enhancement of about one order of magnitude for As and up to three orders of magnitude for Bi. 
To the best of our knowledge, no such calculations have been completed for Sb.
Furthermore, it is unknown how the covalent bonding of the donor to the silicon lattice affects $\gamma_\mathrm{s}$. 
As a result of the uncertainty in $\gamma_\mathrm{s}$, it is difficult to make purely theoretical predictions of $Q$ for donors in silicon.

Recent experiments~\cite{FraPRL15, MorIOPN16, FraPRB16, PlaarXiv16} on quadrupole effects in silicon devices have produced some quantitative results that can be used to estimate $Q$ for As, Sb and Bi donors in Si. 
We will present an analysis for each of the donors in sections below, predicting the quadrupole coupling in the ionized charge state D$^+$, where the EFG is produced by the crystal lattice alone. 

\subsubsection{Arsenic}

While arsenic has the lowest nuclear spin of the donors considered here -- making it less suitable for comparison with classical dynamics -- it is a relatively well-studied donor for its quadrupole properties.
In Refs.~\citenum{FraPRL15} and \citenum{FraPRB16}, spectroscopy of As donors in a strained silicon sample has been performed (uni-axial strain $\epsilon_\perp \approx 3\times 10^{-4}$). 
For an ionized donor, the EFG is generated through the gradient elastic tensor $V = \mathbf{S}\cdot\mathbf{\epsilon}$ (where $\mathbf{\epsilon}$ is the strain tensor in Voigt notation), implying a linear relationship between the applied strain and quadrupole interaction.
Measurement of the quadrupole shifts in two samples of different surface planes [(100) and (111)] enabled the extraction of the nontrivial gradient elastic tensor components, $S_{11} = \SI{1.5e22}{\volt \per \square \meter}$ and $S_{44} = \SI{6.8e22}{\volt \per \square \meter}$. 
These components (which include the Sternheimer antishielding factor $\gamma_\mathrm{s} \approx -7$ for As~\cite{FeiPR69}) can be used to provide a rough estimation of the D$^+$ quadrupole coupling for Bi and Sb (see below).
In a nanodevice, strains of order $10^{-3}$ are expected directly underneath the metallic surface electrodes~\cite{ThoAIPA15} due to the mismatch in thermal expansion coefficients, similar in magnitude to those observed in the prestrained devices in Refs.~\citenum{FraPRL15} and \citenum{FraPRB16}.
This allows us to estimate the quadrupole interaction strength in a device: $Q \approx \SI{210}{\kilo \hertz}$ for a (111) surface and $Q = \SI{60}{\kilo \hertz}$ for a (100) surface.

\subsubsection{Bismuth}

In order to estimate $Q$ in the D$^+$ state for Bi, we require the antishielding factor $\gamma_\mathrm{s}$. 
We take an order-of-magnitude estimate, only serving as a rough guide, of $\gamma_s \approx 100$ for the antishielding factor of Bi.
This value is based on the simulations of measured data reported in Ref. \citenum{MorIOPN16}. 

Using the estimated magnitude of $\gamma_s$ for Bi, the measured gradient elastic tensor components $S_{11}$ and $S_{44}$ of As (converted to strain using the theoretical magnitude of $|\gamma_s| \approx 7$ for As), we estimate the quadrupole interaction strength achievable in a nanodevice for the D$^+$ state to be $Q \approx \SI{240}{\kilo \hertz}$ for a (111) silicon surface and $Q = \SI{45}{\kilo \hertz}$ for a (100) surface.

\subsubsection{Antimony}

Antimony is the least understood of the group-V donors for its quadrupole properties. 
There are no theoretical calculations for the Sternheimer antishielding factor, and no experimental data on the interplay between strain and quadrupole interaction. 
In the EDMR experiments of Ref.~\citenum{MorIOPN16}, $Q$ was found for the neutral $^{121}$Sb donor to be approximately half that of the measured value for $^{209}$B, with the caveat that the implantation conditions were different for the Sb and Bi samples, and no estimate was given for the likely separation between the donors and the readout centers  (which strongly influences the EFG) in the Sb sample.

\section{System characterization via NMR spectroscopy} \label{app: spectroscopy}

High-precision measurements of the nuclear-spin Hamiltonian parameters are needed to achieve an accurate comparison between the quantum system and its classical equivalent.
Furthermore, this is a crucial requirement in enabling arbitrary state preparation as discussed later (Appendix~\ref{app: control}). 
This section describes how the different parameters can be extracted through NMR spectroscopy of the ionized nucleus.

\subsection{NMR spectroscopy} 
Starting from a system where none of the NMR transition frequencies are known, the first step is to find the ESR frequency, which  uniquely depends on the (unknown) nuclear spin state through the hyperfine coupling.
To this end, the electron spin is initialized in the down state through spin-dependent tunneling from the SET island onto the ionized donor, after which a voltage is applied to ensure that the electron is unable to tunnel back.
A pulse with linearly increasing frequency is applied that adiabatically inverts the electron state if its resonance frequency lies within the frequency range~\cite{laucht2014high}.
The bias voltage is then modified to ensure that the electron can only tunnel back onto the SET island if its spin state has successfully been flipped, which is measured as a finite SET current.

Once the first ESR frequency is known, an NMR frequency of the nuclear spin state can be found.
The measurement sequence is nearly identical, with the exception that before loading an electron, a voltage is applied to force the donor into the ionized state, followed by another pulse with linearly increasing frequency in the expected NMR frequency range.
If this pulse sweeps over one of the active NMR frequencies (which have an unique value due to the quadrupole interaction), this will adiabatically invert the nuclear spin state.
As a result, the ESR frequency will change, and so no subsequent adiabatic inversion of the electron spin state will occur.
Repeating the measurement flips the nuclear spin state, resulting in an alternating electron spin flip probability, allowing determination of the NMR frequency. 
The previous two steps can then be repeated to iteratively determine all ESR and NMR frequencies.

The above scheme assumes that loading or unloading an electron leaves the nuclear-spin eigenstates intact, which may require an adiabatic electron-loading scheme (Appendix~\ref{app: control}).

\begin{figure*}[!htb]
\includegraphics[width=\linewidth]{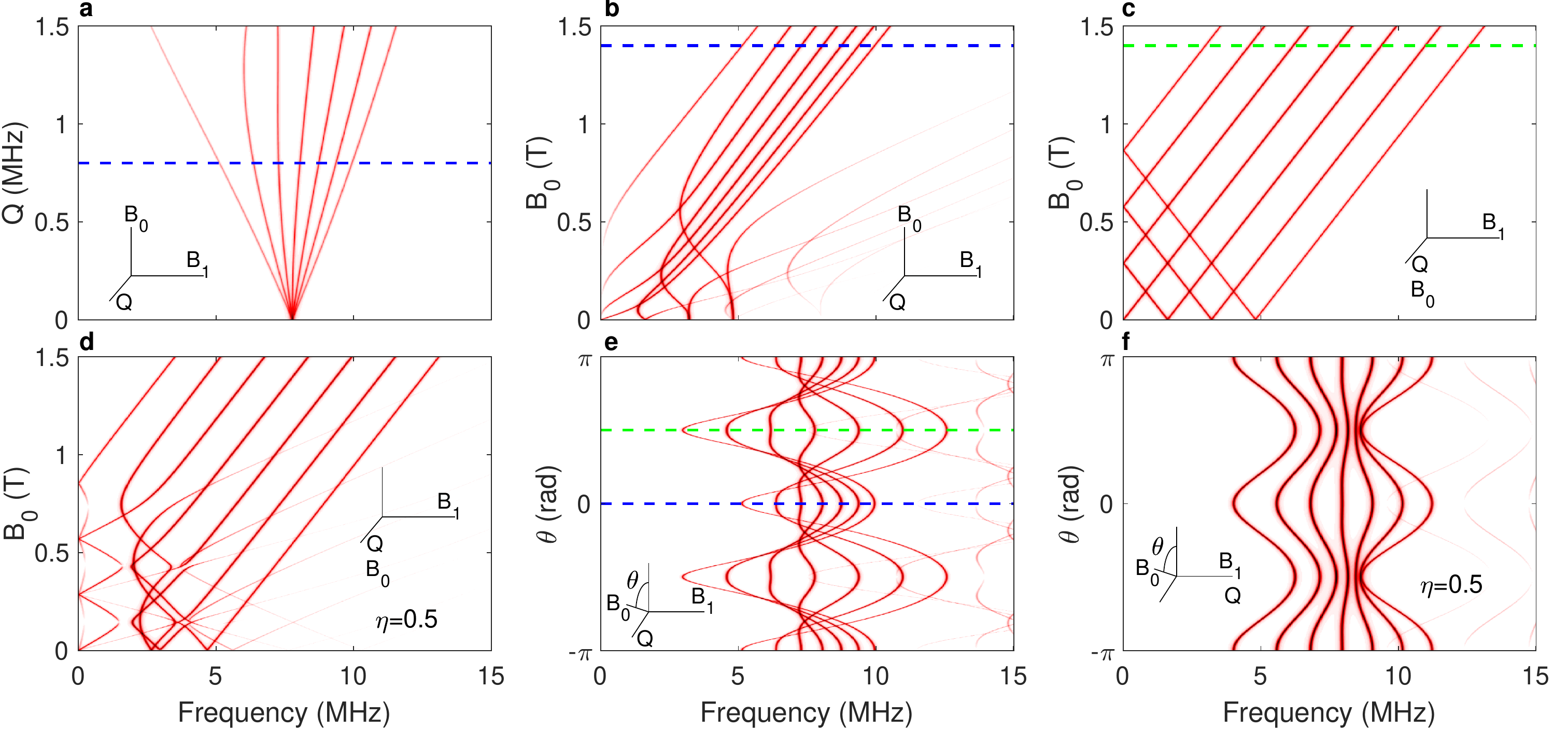}
\caption{
NMR spectra of the ionized \Sb donor ($I=7/2$).  
In all panels, the transition line intensities indicate the coupling between corresponding states, and parameter orientations are given by coordinate axes.
Blue (dark gray) and green (light gray) dashed lines each indicate matching conditions across panels.
Unless specified otherwise, the Hamiltonian is given by Eq.~\eqref{eq: H_donor} of the main text, with $B_0=\SI{1.4}{\tesla}$, $Q=\SI{0.8}{\mega \hertz}$, and $\eta=0$ [blue (dark gray) dashed lines].
(a) Influence of $Q$ on NMR spectra for perpendicular orientations of $B_0$, $Q$, and $B_1$.
The convergence of NMR transition frequencies for $Q=\SI{0}{\hertz}$ shows that a nonzero quadrupole is necessary to distinguish NMR transitions.
(b) Influence of $B_0$ on NMR spectra for perpendicular orientations of $B_0$, $Q$, and $B_1$.
In the absence of a linear interaction ($B_0=\SI{0}{\tesla}$), the twofold degeneracy of the quadratic interaction translates to 3 distinct NMR frequencies.
This degeneracy is lifted when $B_0 > \SI{0}{\tesla}$, and the transition frequencies approach a linear dependence on $B_0$ in the high-field regime.
(c) $B_0$ is varied while oriented along the principal quadrupole axis $\hat{z}'$.
In this case, the eigenstates of the Zeeman interaction are simultaneously eigenstates of the quadrupole interaction, resulting in a linear dependence on $B_0$ and a constant spacing of $2Q$ between successive transitions.
(d) Same as (c) but with $\eta = 0.5$, which results in a nonlinear NMR-frequency dependence on $B_0$. 
The behavior of (c) is recovered in the large-field limit ($Q \ll \gamma_\mathrm{n}B_0$).
(e) Rotating $B_0$ towards principal quadrupole interaction axis, revealing a $\pi$ periodicity and two symmetry axes.
The separation of NMR frequencies is maximal when the orientations of $B_0$ and $Q$ are aligned. 
(f) Rotating $B_0$ perpendicular to $Q$ with $\eta = 0.5$.
The separation between spectral lines is maximal when $B_0$ is aligned with the secondary quadrupole interaction axis $\hat{y}'$. 
When $\eta = 0$, there is no dependence of spectral lines upon rotating $B_0$ perpendicular to $Q$. 
}
\label{fig: figureAppSpectrum}
\end{figure*}

\subsection{Extraction of Hamiltonian parameters}

Combining NMR spectroscopy with full control of both the direction and strength of $B_0$ allows extraction of all Hamiltonian parameters. 
This relies on the underlying assumption  that each of the NMR transitions is individually addressable, a condition that can be satisfied by assuming a $Q$ larger than the NMR transition linewidth [Fig.~\ref{fig: figureAppSpectrum}(a)].
It eases analysis to operate in the large magnetic field limit ($Q\ll \gamma_\mathrm{n} B_0$), where the quadrupole interaction can be treated as a perturbation to Eq.~\eqref{eq: H_donor} of the main text [Fig.~\ref{fig: figureAppSpectrum}(b)].

The first goal is to determine the orientation of the quadrupole coordinate system $(\hat{x}', \hat{y}', \hat{z}')$ [Eq.~\eqref{eq: H_Q} main text].
Once known, aligning the $B_0$-axis with $\hat{z}'$ provides the strength of $Q$, since successive NMR transitions have a constant separation of $2Q$ [Fig.~\ref{fig: figureAppSpectrum}(c)].
This equidistant spacing is lifted by a nonzero asymmetry parameter ($\eta >0$), but is recovered in the high magnetic-field limit [Fig.~\ref{fig: figureAppSpectrum}(d)].

The quadrupole's primary axis $\hat{z}'$ can be found through successive spectroscopy measurements while rotating $B_0$. 
Without prior knowledge of the quadrupole's coordinate system, $B_0$ is initially rotated around an arbitrary axis.
It is guaranteed that such a spectroscopy will reveal two symmetry axes, one of which is perpendicular to $\hat{z}'$.
A second rotation of $B_0$ around this particular symmetry axis will align $B_0$ with $\hat{z}'$ at some specific angle. 
This point has maximum and (nearly) equidistant separation of the spectral lines, thereby revealing the quadrupole's primary axis $\hat{z}'$ [Fig.~\ref{fig: figureAppSpectrum}(e)].

The two remaining unknown parameters of the quadrupole interaction, the asymmetry $\eta$ and its orientation, can be found through a final rotation of $B_0$ around the $\hat{z}'$ axis.
If $\eta = 0$, spectral lines will be independent of this rotation, while for $\eta > 0$ again two symmetry axes will be revealed. 
The symmetry axes with the largest separation of spectral lines corresponds to $B_0$ being parallel to the secondary $\hat{y}'$ axis, and the strength of spectral line variation is determined by the size of $\eta$. 

The sequence sketched here allows variation of a single experimental handle (NMR frequency, $B_0$ direction) to isolate the effect of each parameter. 
This allows an accurate determination of all the relevant Hamiltonian parameters and a detailed understanding of the system.

\section{State preparation and measurement} \label{app: control}

\subsection{Arbitrary state preparation}

\begin{figure}[]
\includegraphics[width=\columnwidth]{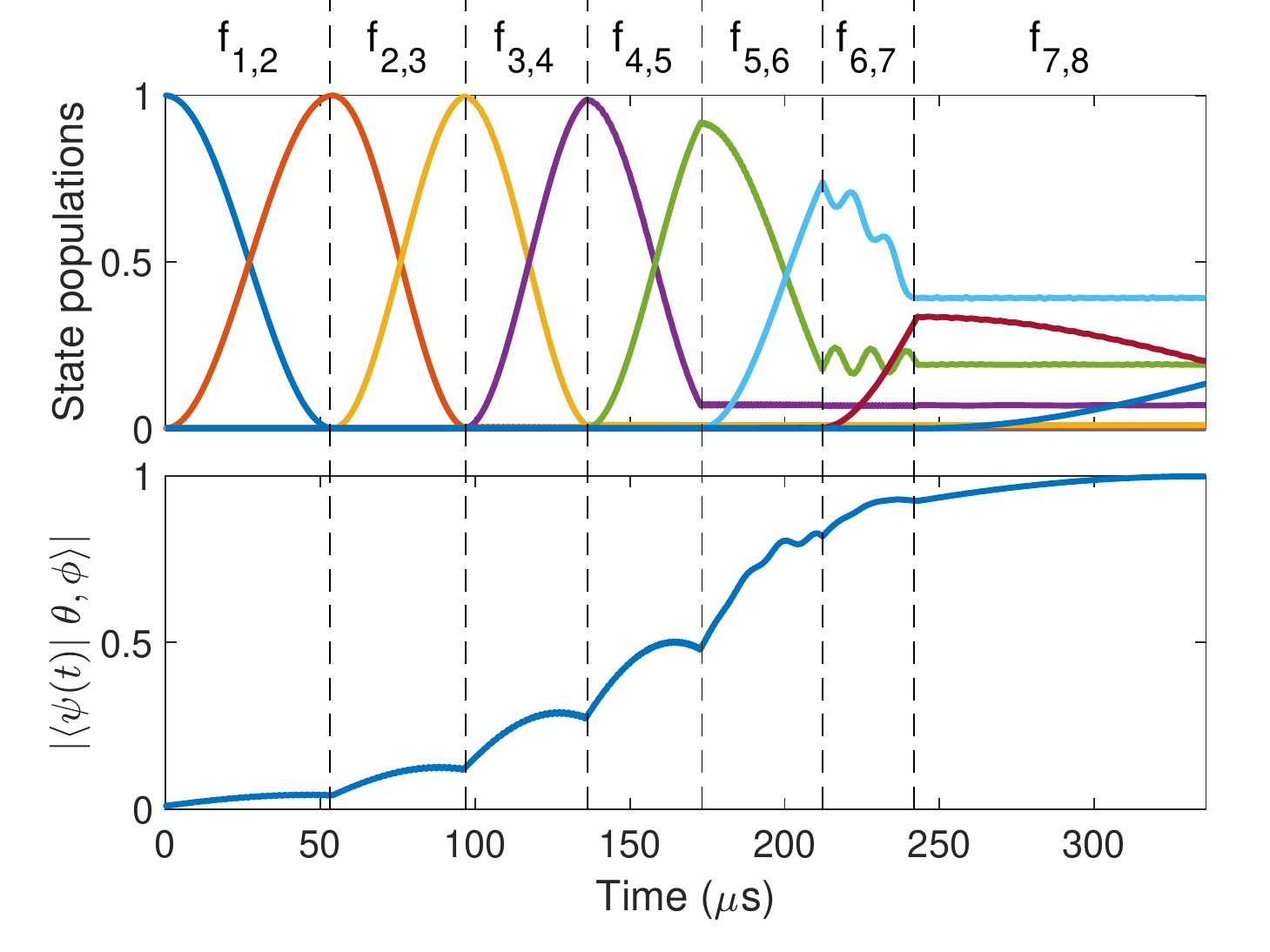}
\caption{Preparation of a spin coherent state.  
Final state is $\ket{\theta, \phi}=\ket{4\pi/5, \pi/2}$, and initial state is the ground state. 
A succession of pulses with different frequencies $f_{k, k'}$ iteratively transfer population from lower-energy eigenstates $\ket{e_k}$ to higher-energy eigenstate $\ket{e_{k'}}$. 
To account for the remaining phase accumulation, intermediate fidelities were calculated after leaving the intermediate state idle until the end of the sequence.
The final fidelity between evolved state and target spin coherent state is $\left|\braket{\psi|\theta, \phi}\right| = 0.9989$ for $B_1= \SI{1}{\milli \tesla}$, $Q=\SI{1}{\mega \hertz}$, $B_0=\SI{0.7}{\tesla}$, and can be further increased by reducing the oscillating magnetic field strength.
}
\label{fig: figureAppStatePreparation}
\end{figure}

The ability to create an arbitrary target state $\ket{\psi_T}$ requires addressability of individual transition frequencies and accurate knowledge of all Hamiltonian parameters. 
The procedure described in this section (see Fig.~\ref{fig: figureAppStatePreparation}) can be used for arbitrary state preparation, provided that the above conditions are fulfilled.

Assuming the system to be initialized in the ground state $\ket{e_1}$, the necessary sequence of pulses that results in the target state $\ket{\psi_T}$ can be found by solving the problem in reverse: go backwards in time and find the pulse sequence to end up in $\ket{e_1}$ starting from $\ket{\psi_T}$.
Starting at $t=\SI{0}{\second}$ with initial state $\ket{\psi_T} = \sum_k a_k \ket{e_k}$, where $a_k = \braket{e_k|\psi_T}$, pulses are iteratively chosen that transfer population from the populated highest-energy eigenstate to a lower-energy eigenstate.
The procedure is as follows:

\begin{enumerate}
\item Choose eigenstate $\ket{e_k}$ with highest eigenvalue $\lambda_k$ and nonzero population $\left|a_k\right|$.
\item Find eigenstate $\ket{e_{k'}}$ with lower eigenvalue $\lambda_{k'}$ that has the highest coupling $\left|\braket{e_{k'}|I_y|e_{k}}\right|$.
\item Transfer the population from $\ket{e_k}$ to $\ket{e_{k'}}$ using a pulse with the following properties:
\begin{itemize}
\item frequency $f_{k, k'}$, which is the transition frequency between $\ket{e_k}$ and $\ket{e_{k'}}$;
\item duration $t_p = \frac{2}{\Omega_{k,k'}} \arctan{\frac{|a_k|}{|a_k'|}}$, where $\Omega_{k, k'}$ is the Rabi frequency;
\item phase is the relative phase difference between $a_k$ and $a_{k'}$, minus the phase offset $2 \pi f_{k, k'} (t-t_p)$.
\end{itemize}
\item Update $t \rightarrow t-t_p$ and repeat steps until all population is transferred to $\ket{e_1}$.
\end{enumerate}

These steps can be followed to create an arbitrary pure state, and in particular the spin coherent states $\ket{\theta, \phi}$ used in the proposed quantum driven-top experiments.
While we do not expect our state-preparation fidelity to be limited by effects such as decoherence (state preparation takes $< \SI{1}{\milli \second}$, $T_2^* \sim \SI{1}{\second}$), the total pulse-sequence duration can be further reduced by using multi frequency pulses.

\subsection{Measurement}

The presence of a strong quadrupole interaction with axis perpendicular to that of the linear interaction inhibits defining a clear quantization axis.
Upon adding an electron to the ionized donor, the accompanying hyperfine interaction can be approximated as an enhancement of the linear interaction by about an order of magnitude for \Sb (two orders for \Bi), significantly altering the eigenbasis of the donor.
The existing techniques for initialization and readout of the nucleus rely on electron tunneling events between donor and SET island.
Here, these tunneling events are accompanied by a change of eigenbasis, resulting in a probabilistically modified state after every such event.
Modified protocols may therefore be necessary for high-fidelity nuclear state initialization and readout.

The proposed solution is the adiabatic transferring of the electron from donor to the silicon/silicon-dioxide interface, which maps neutral-donor eigenstates to ionized-donor eigenstates and vice versa.
This can be achieved by the addition of an electrostatic gate above the donor that can attract its outer electron, a technique that is currently being developed in the context of achieving electrically active transitions and long-range coupling of donor nuclear spin qubits~\cite{tosi2017}.
Once the electron is moved from the donor to the interface, spin-dependent tunneling of the electron from the interface to the SET island allows for readout without affecting the nucleus.

\section{Numerical methods and simulation details} \label{app: methods}

Both the classical and quantum simulations were primarily performed using the commercial software package MATLAB Release 2016b, The MathWorks, Inc., Natick, Massachusetts, United States.
In this section we will describe the simulation techniques for both the classical and the quantum simulations. 
The source code of our simulations is available online in the Supplemental Material \cite{supplemental}.

\subsection{Classical simulations}

The most computationally expensive simulations were those to determine the percentage of phase space that is chaotic [Fig.~\ref{fig: Figure 1}(h),~\ref{fig: Figure 1}(i)). 
Both color maps consist of $25 \times 25$ logarithmically spaced points, each of which corresponds to a particular parameter set.
For each parameter set, a total of 2000 initial angular-momentum coordinates were chosen uniformly distributed over the phase space.
To determine whether the dynamics of an initial coordinate is chaotic, a neighboring point with distance $10^{-8}$ was chosen, and both were evolved for a fixed duration of $100/\alpha$.
Whether or not a trajectory is chaotic is determined by measuring the distance between the two points over time, and fitting this to an exponential curve.
Chaos is characterized by an exponential sensitivity to perturbations, and so the trajectory is categorized as chaotic if its exponent is above a certain threshold. 
This procedure is repeated for each of the 2000 initial conditions, resulting in the percentage of phase space that is chaotic.

Some trajectories, especially near a chaotic-regular boundary, can display an initial exponential divergence but nevertheless behave regularly over sufficient evolution time.
These cases, although uncommon, can result in the trajectory being wrongly categorized as chaotic, and we expect a small uncertainty in the percentages of Fig.~\ref{fig: Figure 1}(h),~\ref{fig: Figure 1}(i). 
It should furthermore be noted that the fitted exponent is not necessarily equal to the Lyapunov exponent, as the intertrajectory distance may have increased sufficiently to be limited by the finite size of the phase space.
Although the exponent is then an underestimate of the Lyapunov exponent, it will certainly be above the chaotic threshold, thereby correctly characterizing the trajectory as chaotic.

The classical simulation results shown in Fig.~\ref{fig: Figure 1} were obtained using the ordinary-differential-equation (ODE) solver SUNDIALS~\cite{hindmarsh2005sundials}.
Communication between MATLAB and Sundials was through self-written C code that was optimized for the driven-top system.
The combination of SUNDIALS and the intermediate C code resulted in a computational speedup of over an order of magnitude compared to the native MATLAB ODE solvers.
As chaotic dynamics are highly sensitive to perturbations, stringent error tolerances were chosen to ensure a high degree of accuracy in the computation of the trajectories.

\subsection{Quantum simulations}

The quantum driven-top system is evolved using the Floquet operator $F$, which can be approximated through segmentation as
\begin{equation}
\mathcal{F} \approx \prod_{k=1}^N e^{-\frac{i t}{\hbar N} \mathcal{H}\left(  \frac{N-k}{N} t \right)},
\end{equation}
which becomes an equality in the limit $N \rightarrow \infty$.
In the simulations, a fixed value of $N=1000$ is used, as results showed that the Floquet operator did not significantly change upon further increasing $N$.
Additionally, the SUNDIALS ODE solver was used to compute $F$, and was found to be nearly identical to $F$ computed using the above method.

Decoherence of spin coherent states under influence of the driven-top Hamiltonian [Eq.~\eqref{eq: H_top_quantum}] was simulated by randomly fluctuating a Hamiltonian parameter during its evolution and averaging over many such evolutions.
To this end, Floquet operators were calculated for 30 values of $Q$ uniformly distributed within three standard deviations of the its mean value ($Q=800 \pm 4$ \si{\kilo \hertz}).
For the evolution, a sequence of Floquet operators was chosen through random sampling of this set using a Gaussian distribution.
This sequence was then applied to all initial spin coherent states, and this process was repeated for 300 such sequences.
For each spin coherent state, the final density matrices were averaged, resulting in a mixed state $\rho$, from which the purity $\Tr(\rho^2)$ was determined.

\end{document}